\documentclass[aps,prd,twocolumn,superscriptaddress]{revtex4}
\usepackage{epsfig,epsf}
\usepackage{amsmath}
\usepackage{amsthm}
\usepackage{amsfonts}
\usepackage{amssymb}
\usepackage{dsfont}
\usepackage{multirow}
\usepackage{appendix}
\usepackage{slashed}
\usepackage[active]{srcltx}
\usepackage{psfrag}

\begin{document}

\title{Properties of the tensor state $bc\overline{b}\overline{c}$}
\date{\today}
\author{S.~S.~Agaev}
\affiliation{Institute for Physical Problems, Baku State University, Az--1148 Baku,
Azerbaijan}
\author{K.~Azizi}
\affiliation{Department of Physics, University of Tehran, North Karegar Avenue, Tehran
14395-547, Iran}
\affiliation{Department of Physics,  Dogus University,  Dudullu-\"{U}mraniye, 34775
Istanbul, T\"{u}rkiye}
\author{H.~Sundu}
\affiliation{Department of Physics Engineering, Istanbul Medeniyet University, 34700
Istanbul, T\"{u}rkiye}

\begin{abstract}
Spectroscopic parameters and decays of the exotic tensor meson $T$ with
content $bc \overline{b}\overline{c}$ are explored in the context of the
diquark-antidiquark model. We treat it as a state built of axial-vector
diquark $b^{T}C\gamma _{\mu }c$ and antidiquark $\overline{b}\gamma _{\nu }C
\overline{c}^{T}$, where $C$ is the charge conjugation matrix. The mass $m$
and current coupling $\Lambda$ of this tetraquark are extracted from
two-point sum rules. Our result for $m=(12.70 \pm 0.09)~\mathrm{GeV}$ proves
that $T$ is unstable against strong dissociations to two-meson final states.
Its dominant decay channels are processes $T \to J/\psi \Upsilon $, $
\eta_{b}\eta _{c}$, and $B_{c}^{(\ast) +}B_{c}^{(\ast) -}$. Kinematically
allowed transformations of $T$ include also decays $T\rightarrow D^{(\ast
)+}D^{(\ast )-}$ and $D^{(\ast )0}\overline{D}^{(\ast )0}$, which are
generated by $b\overline{b}$ annihilation inside of $T$. The full width of $
T $ is estimated by considering all of these channels. Their partial widths
are calculated by invoking methods of three-point sum rule approach which
are required to evaluate strong couplings at corresponding
tetraquark-meson-meson vertices. Our predictions for the mass and width $
\Gamma_{T}=(117.4 \pm 15.9)~ \mathrm{MeV} $ of the tensor state $T$ provide
useful information for experimental studies of fully heavy four-quark exotic
structures.
\end{abstract}

\maketitle


\section{Introduction}

\label{sec:Intro}

During the last years, investigation of fully heavy four-quark mesons has
become one of interesting and rapidly growing branches of high energy
physics. The main reason for such interest, besides pure theoretical
arguments, is observation of four $X$ structures with masses in a range $%
6.2-7.3\ \mathrm{GeV}$ by LHCb-ATLAS-CMS collaborations \cite%
{LHCb:2020bwg,Bouhova-Thacker:2022vnt,CMS:2023owd}. According to
overwhelming opinion they are scalar resonances composed of $cc\overline{c}%
\overline{c}$ quarks, thought there exist kinematical explanations of their
origin as well.

These discoveries generated numerous and interesting publications devoted to
study of newly observed structures \cite%
{Zhang:2020xtb,Albuquerque:2020hio,Yang:2020wkh,Becchi:2020mjz,Becchi:2020uvq,Wang:2022xja, Faustov:2022mvs,Niu:2022vqp,Dong:2022sef,Yu:2022lak,Kuang:2023vac,Wang:2023kir,Dong:2020nwy,Liang:2021fzr}%
. The $X$ resonances were explored also in the QCD sum rule framework in our
articles \cite{Agaev:2023wua,Agaev:2023ruu,Agaev:2023gaq,Agaev:2023rpj}, in
which we modeled them as diquark-antidiquark and hadronic molecule states.
This analysis allowed us to propose our assignments for these resonances.
Thus, some of them were interpreted as a pure ground-level
diquark-antidiquark \cite{Agaev:2023wua} and hadronic molecule \cite%
{Agaev:2023ruu} states, or as admixtures of these two structures \cite%
{Agaev:2023gaq,Agaev:2023rpj}.

Exotic mesons containing only heavy quarks were objects of theoretical
investigations starting from first days of the quark model and quantum
chromodynamics which do not forbid existence of hadrons containing four and
five quarks, pure gluon or quark-gluon systems. Experimental achievements
renewed and intensified interest to these exotic particles. A class of
hidden charm-bottom tetraquarks $bc\overline{b}\overline{c}$ are evidently
among such hadrons. The structures $bc\overline{b}\overline{c}$ were not
discovered yet, but have real chances to be seen in ongoing and future
experiments \cite{Carvalho:2015nqf,Abreu:2023wwg}.

Features of tetraquarks $bc\overline{b}\overline{c}$ with different
spin-parities were considered in the literature \cite%
{Faustov:2022mvs,Wu:2016vtq,Liu:2019zuc,Chen:2019vrj,Bedolla:2019zwg,Cordillo:2020sgc,Weng:2020jao,Yang:2021zrc,Hoffer:2024alv}%
. The masses of tetraquarks $bc\overline{b}\overline{c}$ are the main
parameters calculated in these articles using numerous methods. Information
about partial widths of their decay modes is either scarce or absent. In
other words, our knowledge about properties of exotic mesons $bc\overline{b}%
\overline{c}$ is rather limited. These circumstances, as well as
discrepancies in predictions for the masses made in the different
publications necessitate detailed studies of the tetraquarks $bc\overline{b}%
\overline{c}$.

In Refs.\ \cite{Agaev:2024wvp,Agaev:2024mng}, we investigated the scalar and
axial-vector particles $bc\overline{b}\overline{c}$ and determined their
masses and widths. In the present paper, we extend our analysis by
considering the tensor tetraquark $bc\overline{b}\overline{c}$ with
spin-parity $J^{\mathrm{PC}}=2^{+}$. For simplicity, we label it $T$ and
calculate the mass and full width of this exotic meson. To find the mass $m$
and current coupling $\Lambda $, we use the two-point sum rule (SR) method
\cite{Shifman:1978bx,Shifman:1978by}. Partial widths of numerous decay
channels of $T$ are computed by invoking the three-point sum rule approach.
This is necessary to estimate strong couplings at relevant
tetraquark-meson-meson vertices which determine widths of the processes
under analysis.

There are a few types of decay modes of the tetraquark $T$. Decays to pairs
of quarkonia $J/\psi \Upsilon $ and $\eta _{b}\eta _{c}$, as well as
processes $T\rightarrow B_{c}^{\ast +}B_{c}^{\ast -}$ and $%
B_{c}^{+}B_{c}^{-} $ are dissociations of the initial particle to
final-state mesons. These decays are dominant channels of $T$ in which four
constituent quarks form the final-state conventional mesons. The second kind
of processes is triggered by annihilation in $T$ of $b\overline{b}$ quarks
to a pair of light quarks and subsequent generation of $DD$ mesons with
suitable electric charges and spin-parities. In the case of the tensor
tetraquark, we limit ourselves by investigation of four decays $T\rightarrow
D^{(\ast )+}D^{(\ast )-}$ and $D^{(\ast )0}\overline{D}^{(\ast )0}$.

This work is composed of the following parts: In Sec.\ \ref{sec:Mass}, we
calculate the mass and current coupling of the tensor state $T$. Partial
widths of decays $T\rightarrow J/\psi \Upsilon $ and $\eta _{b}\eta _{c}$
are computed in Sec.\ \ref{sec:Widths1}. The processes with $B_{c}^{(\ast
)+}B_{c}^{(\ast )-}$ mesons in final states are considered in the next Sec. %
\ref{sec:Widths2}. Partial widths of the decays $T\rightarrow D^{(\ast
)+}D^{(\ast )-}$ and $D^{(\ast )0}\overline{D}^{(\ast )0}$ are evaluated in
Sec.\ \ref{sec:Widths3}. In this section, we also find the full width of the
tensor tetraquark $T$. We make our conclusions in the last part of the paper
Sec. \ref{sec:Conc}.


\section{Mass $m$ and current coupling $\Lambda $ of the tetraquark $T$}

\label{sec:Mass}
Spectroscopic parameters of the tetraquark $T$ are quantities that
characterize this particle and determine its possible decay modes. The mass $%
m$ and current coupling $\Lambda $ of a particle can be evaluated using
different approaches. One of the effective nonperturbative tools to find
these parameters is the two-point sum rule method \cite%
{Shifman:1978bx,Shifman:1978by}. Originally invented to study parameters of
the ordinary baryons and mesons, it can be successfully applied for analysis
of exotic hadrons as well \cite{Albuquerque:2018jkn,Agaev:2020zad}.

In the framework of this method one has to extract SRs for $m$ and $\Lambda $%
, which can be done by considering the correlation function
\begin{equation}
\Pi _{\mu \nu \alpha \beta }(p)=i\int d^{4}xe^{ipx}\langle 0|\mathcal{T}%
\{J_{\mu \nu }(x)J_{\alpha \beta }^{\dag }(0)\}|0\rangle ,  \label{eq:CF1}
\end{equation}%
where $J_{\mu \nu }(x)$ is the interpolating current for the tensor
tetraquark and $\mathcal{T}$ \ is the time-ordered product of two currents.

Analytical expression of $J_{\mu \nu }(x)$ depends on a diquark-antidiquark
model chosen for the particle. In the present article, we consider $T$ as a
diquark-antidiquark state composed of an axial-vector diquark $b^{T}C\gamma
_{\mu }c$ and antidiquark $\overline{b}\gamma _{\nu }C\overline{c}^{T}$.
Accordingly, the interpolating current $J_{\mu \nu }(x)$ has the following
form
\begin{eqnarray}
J_{\mu \nu }(x) &=&b_{a}^{T}(x)C\gamma _{\mu }c_{b}(x)\left[ \overline{b}%
_{a}(x)\gamma _{\nu }C\overline{c}_{b}^{T}(x)\right.  \notag \\
&&\left. -\overline{b}_{b}(x)\gamma _{\nu }C\overline{c}_{a}^{T}(x)\right] .
\label{eq:CR1}
\end{eqnarray}%
Here, $C$ is the charge conjugation matrix, whereas $a$ and $b$ are the
color indices. The current $J_{\mu \nu }$ describes the tetraquark with
spin-parities $J^{\mathrm{P}}=2^{+}$.

To find the sum rules for the mass $m$ and current coupling $\Lambda $, we
first have to compute the correlation function $\Pi _{\mu \nu \alpha \beta
}(p)$ using physical parameters of the tetraquark. For these purposes, we
insert into Eq.\ (\ref{eq:CF1}) a full set of states with the quark content
and spin-parities of the tetraquark $T$, and integrate it over the variable $%
x$. Then the correlator becomes equal to
\begin{equation}
\Pi _{\mu \nu \alpha \beta }^{\mathrm{Phys}}(p)=\frac{\langle 0|J_{\mu \nu
}|T(p,\epsilon )\rangle \langle T(p,\epsilon )|J_{\alpha \beta }^{\dag
}|0\rangle }{m^{2}-p^{2}}+\cdots,  \label{eq:Phys1}
\end{equation}%
where the term in Eq.\ (\ref{eq:Phys1}) is the contribution of the
ground-state particle $T$, whereas the dots show contributions of higher
resonances and continuum states. Here, $\epsilon =\epsilon _{\mu \nu
}^{(\lambda )}(p)$ is the polarization tensor of the tetraquark $T$. For
further calculations, it is convenient to introduce the matrix element
\begin{equation}
\langle 0|J_{\mu \nu }|T(p,\epsilon (p)\rangle =\Lambda \epsilon _{\mu \nu
}^{(\lambda )}(p).  \label{eq:ME1}
\end{equation}%
To find $\Pi _{\mu \nu \alpha \beta }^{\mathrm{Phys}}(p)$ we substitute Eq. (%
\ref{eq:ME1})\ into the correlator Eq. (\ref{eq:Phys1}) and perform
summation over polarization tensor using
\begin{eqnarray}
\sum\limits_{\lambda }\epsilon _{\mu \nu }^{(\lambda )}(p)\epsilon _{\alpha
\beta }^{\ast (\lambda )}(p) &=&\frac{1}{2}(\widetilde{g}_{\mu \alpha }%
\widetilde{g}_{\nu \beta }+\widetilde{g}_{\mu \beta }\widetilde{g}_{\nu
\alpha })  \notag \\
&&-\frac{1}{3}\widetilde{g}_{\mu \nu }\widetilde{g}_{\alpha \beta },
\label{eq:F1}
\end{eqnarray}%
where%
\begin{equation}
\widetilde{g}_{\mu \nu }=-g_{\mu \nu }+\frac{p_{\mu }p_{\nu }}{p^{2}}.
\label{eq:F2}
\end{equation}%
Our computations yield
\begin{eqnarray}
\Pi _{\mu \nu \alpha \beta }^{\mathrm{Phys}}(p) &=&\frac{\Lambda ^{2}}{%
m^{2}-p^{2}}\left\{ \frac{1}{2}\left( g_{\mu \alpha }g_{\nu \beta }+g_{\mu
\beta }g_{\nu \alpha }\right) \right.  \notag \\
&&\left. +\text{other structures}\right\} +..,  \label{eq:Phys2}
\end{eqnarray}%
with ellipses standing for contributions of other structures as well as
higher resonances and continuum states. Note that, after application of
Eqs.\ (\ref{eq:F1}) and (\ref{eq:F2}) there appear numerous Lorentz
structures in the curly brackets. The term proportional to $(g_{\mu \alpha
}g_{\nu \beta }+g_{\mu \beta }g_{\nu \alpha })$ contains contribution of
only spin-$2$ particle, whereas remaining components in Eq.\ (\ref{eq:Phys2}%
) are formed due to contributions of spin-$0$ and -$1$ states as well.
Therefore, in our studies we restrict ourselves by exploring this term and
corresponding invariant amplitude $\Pi ^{\mathrm{Phys}}(p^{2})$.

At next phase of investigations, we compute the correlator $\Pi _{\mu \nu
\alpha \beta }(p)$ with some accuracy in the operator product expansion ($%
\mathrm{OPE}$). To this end, we have to insert the explicit expression of
the current $J_{\mu \nu }(x)$ into Eq.\ (\ref{eq:CF1}) and contract relevant
quark fields to obtain $\Pi _{\mu \nu \alpha \beta }^{\mathrm{OPE}}(p)$. As
a result, we find
\begin{eqnarray}
&&\Pi _{\mu \nu \alpha \beta }^{\mathrm{OPE}}(p)=i\int d^{4}xe^{ipx}\mathrm{%
Tr}\left[ \gamma _{\alpha }\widetilde{S}_{b}^{aa^{\prime }}(x)\gamma _{\mu
}S_{c}^{bb^{\prime }}(x)\right]  \notag \\
&&\times \left\{ \mathrm{Tr}\left[ \gamma _{\nu }\widetilde{S}%
_{c}^{b^{\prime }b}(-x)\gamma _{\beta }S_{b}^{a^{\prime }a}(-x)\right] -%
\mathrm{Tr}\left[ \gamma _{\nu }\widetilde{S}_{c}^{a^{\prime }b}(-x)\right.
\right. ,  \notag \\
&&\left. \times \gamma _{\beta }S_{b}^{b^{\prime }a}(-x)\right] +\mathrm{Tr}%
\left[ \gamma _{\nu }\widetilde{S}_{c}^{a^{\prime }a}(-x)\gamma _{\beta
}S_{b}^{b^{\prime }b}(-x)\right]  \notag \\
&&\left. -\mathrm{Tr}\left[ \gamma _{\nu }\widetilde{S}_{c}^{b^{\prime
}a}(-x)\gamma _{\beta }S_{b}^{a^{\prime }b}(-x)\right] \right\} ,
\label{eq:QCD1}
\end{eqnarray}%
where
\begin{equation}
\widetilde{S}_{b(c)}(x)=CS_{b(c)}^{T}(x)C,  \label{eq:Prop}
\end{equation}%
and $S_{b(c)}(x)$ are $b$ and $c$ quarks' propagators.

The massive quark propagator $S_{Q}(x)$ was calculated using an external
field method at the fixed-point gauge (for details, see Ref.\ \cite%
{Reinders:1984sr}). More recent expression for $S_{Q}(x)$ can be found in
Ref.\ \cite{Agaev:2020zad} which contains terms $\sim g_{s}^{3}G^{3}$. It
depends only on gluon fields, as a result, the correlator $\Pi _{\mu \nu
\alpha \beta }^{\mathrm{OPE}}(p)$ contains merely gluon vacuum condensates.
In our calculations we take into account nonperturbative contributions $\sim
\langle \alpha _{s}G^{2}/\pi \rangle $, therefore adopt the following
expression for the propagator $S_{Q}(x)$
\begin{eqnarray}
S_{Q}^{ab}(x) &=&i\int \frac{d^{4}k}{(2\pi )^{4}}e^{-ikx}\Bigg \{\frac{%
\delta _{ab}\left( {\slashed k}+m_{Q}\right) }{k^{2}-m_{Q}^{2}}  \notag \\
&&-\frac{g_{s}G_{ab}^{\alpha \beta }}{4}\frac{\sigma _{\alpha \beta }\left( {%
\slashed k}+m_{Q}\right) +\left( {\slashed k}+m_{Q}\right) \sigma _{\alpha
\beta }}{(k^{2}-m_{Q}^{2})^{2}}  \notag \\
&&+\frac{g_{s}^{2}G^{2}}{12}\delta _{ab}m_{Q}\frac{k^{2}+m_{Q}{\slashed k}}{%
(k^{2}-m_{Q}^{2})^{4}}+\cdots \Bigg \}.  \label{eq:QProp}
\end{eqnarray}%
Here, we have introduced the notations
\begin{equation}
G_{ab}^{\alpha \beta }\equiv G_{A}^{\alpha \beta }\lambda _{ab}^{A}/2,\ \
G^{2}=G_{\alpha \beta }^{A}G_{A}^{\alpha \beta },\ A=1-8,
\end{equation}%
with $G_{A}^{\alpha \beta }$ being the gluon field-strength tensor, and $%
\lambda ^{A}$--Gell-Mann matrices.

Having extracted the structure $(g_{\mu \alpha }g_{\nu \beta }+g_{\mu \beta
}g_{\nu \alpha })$ from $\Pi _{\mu \nu \alpha \beta }^{\mathrm{OPE}}(p)$ and
labeled corresponding invariant amplitude by $\Pi ^{\mathrm{OPE}}(p^{2})$,
one can derive SRs for the mass and current coupling of the tetraquark $T$.
In fact, the function $\Pi ^{\mathrm{Phys}}(p^{2})$ can be expressed as the
dispersion integral%
\begin{equation}
\Pi ^{\mathrm{Phys}}(p^{2})=\int_{4\mathcal{M}^{2}}^{\infty }\frac{\rho ^{%
\mathrm{Phys}}(s)ds}{s-p^{2}}+\cdots ,  \label{eq:DisRel}
\end{equation}%
where $\mathcal{M}^{2}=(m_{b}+m_{c})^{2}$ and the dots indicate subtraction
terms required to render finite $\Pi ^{\mathrm{Phys}}(p^{2})$. The spectral
density $\rho ^{\mathrm{Phys}}(s)$ is equal to the imaginary part of $\Pi ^{%
\mathrm{Phys}}(p^{2})$,%
\begin{equation}
\rho ^{\mathrm{Phys}}(s)=\Lambda ^{2}\delta (s-m^{2})+\rho ^{\mathrm{h}%
}(s)\theta (s-s_{0}).  \label{eq:SDensity}
\end{equation}%
Here, $\theta (z)$ is the unit step function, and $s_{0}$ is the continuum
subtraction parameter. The contribution of the ground-level particle in Eq.\
(\ref{eq:SDensity}) is separated from other effects and represented by the
pole term. Contributions to $\rho ^{\mathrm{Phys}}(s)$ coming from higher
resonances and continuum states are characterized by an unknown hadronic
spectral density $\rho ^{\mathrm{h}}(s)$. It is clear that $\rho ^{\mathrm{%
Phys}}(s)$ leads to the expression
\begin{equation}
\Pi ^{\mathrm{Phys}}(p^{2})=\frac{\Lambda ^{2}}{m^{2}-p^{2}}%
+\int_{s_{0}}^{\infty }\frac{\rho ^{\mathrm{h}}(s)ds}{s-p^{2}}.
\label{eq:InvAmp2}
\end{equation}

Theoretically, the amplitude $\Pi ^{\mathrm{OPE}}(p^{2})$ can be calculated
in deep Euclidean region $p^{2}\ll 0$ using the operator product expansion.
The coefficient functions in $\mathrm{OPE}$ could be obtained using methods
of perturbative QCD, whereas nonperturbative information is contained in the
gluon condensate $\langle \alpha _{s}G^{2}/\pi \rangle $.

Having continued $\Pi ^{\mathrm{OPE}}(p^{2})$ analytically to the Minkowski
domain and found its imaginary part, we get the two-point spectral density $%
\rho ^{\mathrm{OPE}}(s)$. In the region $p^{2}\ll 0$ we apply the Borel
transformation $\mathcal{B}$ to remove subtraction terms in the dispersion
integral and suppress contributions of higher resonances and continuum
states. For $\mathcal{B}\Pi ^{\mathrm{Phys}}(p^{2})$, we obtain%
\begin{equation}
\mathcal{B}\Pi ^{\mathrm{Phys}}(p^{2})=\Lambda
^{2}e^{-m^{2}/M^{2}}+\int_{s_{0}}^{\infty }ds\rho ^{\mathrm{h}%
}(s)e^{-s/M^{2}},  \label{eq:CorBor}
\end{equation}%
where $M^{2}$ is the Borel parameter. One can write the dispersion
representation for the amplitude $\Pi ^{\mathrm{OPE}}(p^{2})$ using $\rho ^{%
\mathrm{OPE}}(s)$ as well. Then, by equating the Borel transformations of $%
\Pi ^{\mathrm{Phys}}(p^{2})$ and $\Pi ^{\mathrm{OPE}}(p^{2})$ and applying
the assumption about hadron-parton duality $\rho ^{\mathrm{h}}(s)\simeq \rho
^{\mathrm{OPE}}(s)$ in a duality region, we subtract the second term in Eq.\
(\ref{eq:CorBor}) from the QCD side of the obtained equality and get
\begin{equation}
\Lambda ^{2}e^{-m^{2}/M^{2}}=\Pi (M^{2},s_{0}).  \label{eq:SR}
\end{equation}%
Here,
\begin{equation}
\Pi (M^{2},s_{0})=\int_{4\mathcal{M}^{2}}^{s_{0}}ds\rho ^{\mathrm{OPE}%
}(s)e^{-s/M^{2}}+\Pi (M^{2}).  \label{eq:CorrF}
\end{equation}%
The nonperturbative function $\Pi (M^{2})$ is computed directly from the
correlator $\Pi ^{\mathrm{OPE}}(p)$ and contains contributions that do not
enter to the spectral density.

After simple manipulations, we get
\begin{equation}
m^{2}=\frac{\Pi ^{\prime }(M^{2},s_{0})}{\Pi (M^{2},s_{0})},  \label{eq:Mass}
\end{equation}%
and
\begin{equation}
\Lambda ^{2}=e^{m^{2}/M^{2}}\Pi (M^{2},s_{0}),  \label{eq:Coupl}
\end{equation}%
which are the sum rules for $m$ and $\Lambda $, respectively. In Eq.\ (\ref%
{eq:Mass}), we also use the short-hand notation $\Pi ^{\prime
}(M^{2},s_{0})=d\Pi (M^{2},s_{0})/d(-1/M^{2})$. The spectral density $\rho ^{%
\mathrm{OPE}}(s)$ contains the perturbative $\rho ^{\mathrm{pert.}}(s)$ and
nonperturbative $\rho ^{\mathrm{Dim4}}(s)$ terms. Explicit expressions for $%
\rho ^{\mathrm{pert.}}(s)$ and $\Pi (M^{2})$ are presented in the Appendix.

We need to specify the input parameters in Eqs.\ (\ref{eq:Mass}) and (\ref%
{eq:Coupl}) to perform numerical computations. Some of them are universal
quantities and do not depend on a problem under consideration. The masses of
$b$ and $c$ quarks and gluon vacuum condensate $\langle \alpha _{s}G^{2}/\pi
\rangle $ are such parameters. In the present work, we use the following
values:
\begin{eqnarray}
&&m_{b}=4.18_{-0.02}^{+0.03}~\mathrm{GeV},\ m_{c}=(1.27\pm 0.02)~\mathrm{GeV}%
,  \notag \\
&&\langle \alpha _{s}G^{2}/\pi \rangle =(0.012\pm 0.004)~\mathrm{GeV}^{4}.
\label{eq:GluonCond}
\end{eqnarray}%
The $m_{b}$ and $m_{c}$ are the running quark masses in the $\overline{%
\mathrm{MS}}$ scheme \cite{PDG:2022}. The gluon vacuum condensate was
extracted from analysis of various hadronic processes in Refs.\ \cite%
{Shifman:1978bx,Shifman:1978by}.

\begin{figure}[h]
\includegraphics[width=8.5cm]{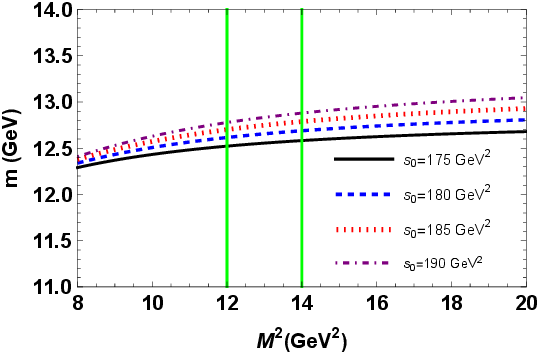}
\caption{Mass $m$ of the tetraquark $T$ as a function of $M^{2}$ for fixed $%
s_{0}$. The two vertical lines fix values of the Borel parameter, where $m$
is extracted. }
\label{fig:WideRange}
\end{figure}
Contrary, the Borel and continuum subtraction parameters $M^{2}$ and $s_{0}$
are specific for each problem and should satisfy some standard constraints
of SR computations. Dominance of the pole contribution ($\mathrm{PC}$) in
extracted quantities and their stability upon variations of $M^{2}$ and $%
s_{0}$ as well as convergence of the operator product expansion are
important conditions for correct SR analysis. To fulfill these requirements,
we impose on the parameters $M^{2}$ and $s_{0}$ the following restrictions.
First, the pole contribution
\begin{equation}
\mathrm{PC}=\frac{\Pi (M^{2},s_{0})}{\Pi (M^{2},\infty )},  \label{eq:PC}
\end{equation}%
should obey $\mathrm{PC}\geq 0.5$. The convergence of $\mathrm{OPE}$ is
second important condition in the SR analysis. Because the correlation
function contains only the nonperturbative dimension-$4$ term $\Pi ^{\mathrm{%
Dim4}}(M^{2},s_{0})$, we require fulfilment of the constraint $|\Pi ^{%
\mathrm{Dim4}}(M^{2},s_{0})|=0.05\Pi (M^{2},s_{0})$, which ensures the
convergence of the operator product expansion. It is worth noting that the
maximum of the Borel parameter is determined from Eq.\ (\ref{eq:PC}),
whereas convergence of $\mathrm{OPE}$ allows us to fix its minimal value.

\begin{figure}[h]
\includegraphics[width=8.5cm]{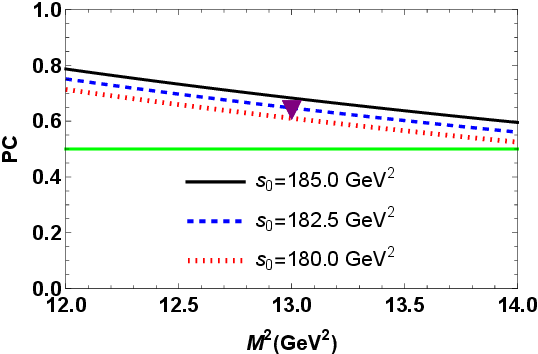}
\caption{Dependence of $\mathrm{PC}$ on the Borel parameter $M^{2}$ at fixed
$s_{0}$. The horizontal green line corresponds to $\mathrm{PC}=0.5$. The red
triangle shows the point $M^{2}=13~\mathrm{GeV}^{2}$ and $s_{0}=182.5~%
\mathrm{GeV}^{2}$. }
\label{fig:PC}
\end{figure}

Numerical calculations are performed over a wide range of the parameters $%
M^{2}$ and $s_{0}$. In Fig.\ \ref{fig:WideRange}, we plot the mass $m$ in
the range $M^{2}=8-20~\mathrm{GeV}^{2}$ at some fixed $s_{0}$. Analysis of
these results allows us to fix the working windows for $M^{2}$ and $s_{0}$,
where all aforementioned restrictions are obeyed. We find that the regions
\begin{equation}
M^{2}\in \lbrack 12,14]~\mathrm{GeV}^{2},\ s_{0}\in \lbrack 180,185]~\mathrm{%
GeV}^{2}  \label{eq:Wind1}
\end{equation}%
comply with these constraints. Indeed, on the average in $s_{0}$ at maximal
and minimal $M^{2}$ the pole contribution is $\mathrm{PC}\approx 0.56$ and $%
\mathrm{PC}\approx 0.75$, respectively. The nonperturbative term is positive
and at $M^{2}=12~\mathrm{GeV}^{2}$ forms less than $1\%$ of the whole
result. The dependence of $\mathrm{PC}$ on the Borel parameter is plotted in
Fig.\ \ref{fig:PC}, in which all curves exceed the limit line $\mathrm{PC}%
=0.5$.

\begin{widetext}

\begin{figure}[h!]
\begin{center}
\includegraphics[totalheight=6cm,width=8cm]{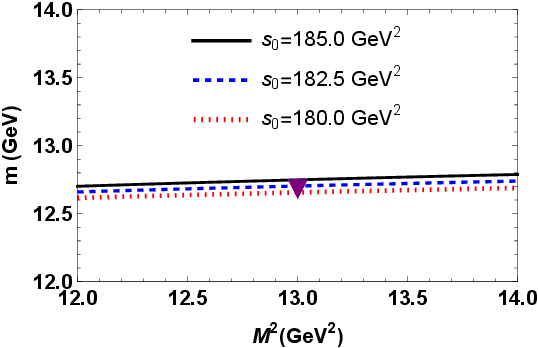}
\includegraphics[totalheight=6cm,width=8cm]{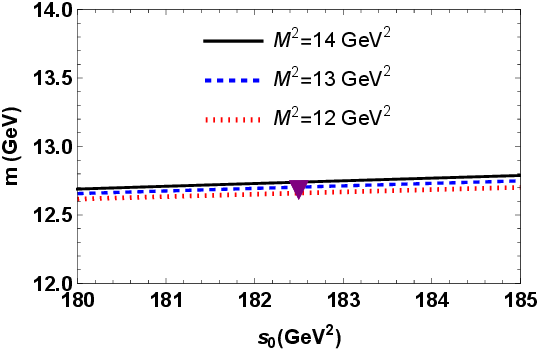}
\end{center}
\caption{Mass $m$ as a function of the Borel  $M^{2}$ (left panel), and continuum threshold $s_0$ parameters (right panel).}
\label{fig:Mass}
\end{figure}

\end{widetext}

To extract $m$ and $\Lambda $, we compute their mean values over the regions
Eq.\ (\ref{eq:Wind1}) and find
\begin{eqnarray}
m &=&(12.70\pm 0.09)~\mathrm{GeV},  \notag \\
\Lambda &=&(2.16\pm 0.24)~\mathrm{GeV}^{5}.  \label{eq:Result1}
\end{eqnarray}%
Effectively, results in Eq.\ (\ref{eq:Result1}) are equal to SR predictions
at the point $M^{2}=13~\mathrm{GeV}^{2}$ and $s_{0}=182.5~\mathrm{GeV}^{2}$,
where $\mathrm{PC}\approx 0.65$, which guarantees the dominance of $\mathrm{%
PC}$ in the extracted parameters. Uncertainties in Eq.\ (\ref{eq:Result1})
are generated mainly by the choices of $M^{2}$ and $s_{0}$. These
theoretical errors form only $\pm 0.7\%$ of the mass $m$, which demonstrates
the high stability of the obtained prediction. Such accuracy of the result
is connected with the SR for $m$, Eq.\ (\ref{eq:Mass}), which determines it
as a ratio of the correlation functions. Therefore, changes in the
correlators due to $M^{2}$, and $s_{0}$ compensate each other in $m$ and
stabilize in this way the numerical output. In the case of $\Lambda $ errors
amount to $\pm 11\%$ of the central value, but still remain within limits
acceptable for the sum rule analysis. In Fig.\ \ref{fig:Mass}, we show $m$
as a function of $M^{2}$ and $s_{0}$.

The mass of the tensor tetraquark $T$ was evaluated in the framework of
different models and methods \cite%
{Faustov:2022mvs,Wu:2016vtq,Liu:2019zuc,Chen:2019vrj,Bedolla:2019zwg,Cordillo:2020sgc,Weng:2020jao,Yang:2021zrc}%
. In the relativistic quark model the authors obtained $12.849\ \mathrm{GeV}$
\cite{Faustov:2022mvs}. A considerably larger result, i.e., $13.59-13.599~%
\mathrm{GeV}$ was found in the color-magnetic interaction model \cite%
{Wu:2016vtq}. The mass spectra of all-heavy tetraquarks with different
contents were investigated in Ref.\ \cite{Liu:2019zuc}, in which for the
tensor state $bc\overline{b}\overline{c}$ \ the authors got $12.993-13.021~%
\mathrm{GeV}$. The nonrelativistic chiral quark model led to $12.809~\mathrm{%
GeV}$ \cite{Chen:2019vrj}. In the relativized diquark Hamiltonian model the
mass of the tensor tetraquark $2^{++}$ depending on diquarks' spins and
total spin and orbital angular momentum of the tetraquark changes from $%
12.576~\mathrm{GeV}$ to $13.65~\mathrm{GeV}$ \cite{Bedolla:2019zwg}$\mathrm{.%
}$ Prediction $12.582~\mathrm{GeV}$ for $m$ was made in Ref. \cite%
{Cordillo:2020sgc}. In the extended chromomagnetic model \cite{Weng:2020jao}
this tensor has the mass in the range $12.537$ and $12.754~\mathrm{GeV}$.
This problem was addressed also in the SR framework \cite{Yang:2021zrc}.
Values for the mass of the tensor tetraquarks $bc\overline{b}\overline{c}$
modeled by a color antitriplet-triplet and sextet-antisextet interpolating
currents are equal to $12.30$ and $12.35~\mathrm{GeV}$, respectively.

\section{Decays $T\rightarrow J/\protect\psi \Upsilon $ and $T\rightarrow
\protect\eta _{b}\protect\eta _{c}$}

\label{sec:Widths1}

Information on the mass of the tensor state $T$ permits us to make
conclusions about its decay channels. Decays to quarkonium pairs $J/\psi
\Upsilon $ and $\eta _{b}\eta _{c}$ are among kinematically possible decay
modes of $T$. Indeed, thresholds for creation of these final states are $%
12.558~\mathrm{GeV}$ and $12.383~\mathrm{GeV}$, respectively. In this
section we study these decay channels of $T$ .


\subsection{Process $T\rightarrow J/\protect\psi \Upsilon $}


Here, we consider the decay $T\rightarrow J/\psi \Upsilon $, the partial
width of which, apart from usual input parameters, is determined by the
strong coupling $g_{1}$ at the vertex $TJ/\psi \Upsilon $. The coupling $%
g_{1}$ can be evaluated using the form factor $g_{1}(q^{2})$ at the mass
shell $q^{2}=m_{J/\psi }^{2}$.

We are going to derive the three-point sum rule for the form factor $%
g_{1}(q^{2})$ from analysis of the correlation function
\begin{eqnarray}
\Pi _{\mu \nu \alpha \beta }(p,p^{\prime }) &=&i^{2}\int
d^{4}xd^{4}ye^{ip^{\prime }y}e^{-ipx}\langle 0|\mathcal{T}\{J_{\mu
}^{\Upsilon }(y)  \notag \\
&&\times J_{\nu }^{J/\psi }(0)J_{\alpha \beta }^{\dagger }(x)\}|0\rangle ,
\label{eq:CF1a}
\end{eqnarray}%
where $J_{\mu }^{\Upsilon }(x)$ and $J_{\nu }^{J/\psi }(x)$ are
interpolating currents of the vector quarkonia $\Upsilon $ and $J/\psi $,
respectively. They are defined as
\begin{equation}
J_{\mu }^{\Upsilon }(x)=\overline{b}_{i}(x)\gamma _{\mu }b_{i}(x),\ J_{\nu
}^{J/\psi }(x)=\overline{c}_{j}(x)\gamma _{\nu }c_{j}(x),
\end{equation}%
with $i$ and $j$ being the color indices.

To find the physical side of the sum rule $\Pi _{\mu \nu \alpha \beta }^{%
\mathrm{Phys}}(p,p^{\prime })$, we need to rewrite Eq.\ (\ref{eq:CF1a})
using the involved particles' physical parameters. By taking into account
only contributions of the ground-level states, we recast the correlator $\Pi
_{\mu \nu \alpha \beta }(p,p^{\prime })$ into the form
\begin{eqnarray}
&&\Pi _{\mu \nu \alpha \beta }^{\mathrm{Phys}}(p,p^{\prime })=\frac{\langle
0|J_{\mu }^{\Upsilon }|\Upsilon (p^{\prime },\varepsilon _{1})\rangle }{%
p^{\prime 2}-m_{\Upsilon }^{2}}\frac{\langle 0|J_{\nu }^{J/\psi }|J/\psi
(q,\varepsilon _{2})\rangle }{q^{2}-m_{J/\psi }^{2}}  \notag \\
&&\times \langle \Upsilon (p^{\prime },\varepsilon _{1})J/\psi
(q,\varepsilon _{2})|T(p,\epsilon )\rangle \frac{\langle T(p,\varepsilon
)|J_{\alpha \beta }^{\dagger }|0\rangle }{p^{2}-m^{2}}+\cdots ,  \notag \\
&&  \label{eq:TP1}
\end{eqnarray}%
where $m_{\Upsilon }=(9460.40\pm 0.09\pm 0.04)~\mathrm{MeV}$ and $m_{J/\psi
}=(3096.900\pm 0.006)~\mathrm{MeV}$ are masses of the $\Upsilon $ and $%
J/\psi $ mesons \cite{PDG:2022}. In the expression above, we denote by $%
\varepsilon _{1}$ and $\varepsilon _{2}$ the polarization vectors of these
quarkonia, respectively.

To further simplify Eq.\ (\ref{eq:TP1}), it is convenient to employ the
matrix elements of the mesons $\Upsilon $ and $J/\psi $%
\begin{eqnarray}
\langle 0|J_{\mu }^{\Upsilon }|\Upsilon (p^{\prime },\varepsilon
_{1})\rangle &=&f_{\Upsilon }m_{\Upsilon }\varepsilon _{1\mu }(p^{\prime }),
\notag \\
\langle 0|J_{\nu }^{J/\psi }|J/\psi (q,\varepsilon _{2})\rangle &=&f_{J/\psi
}m_{J/\psi }\varepsilon _{2\nu }(q).  \label{eq:C2}
\end{eqnarray}%
Here, $f_{\Upsilon }=(708\pm 8)~\mathrm{MeV}$ and $f_{J/\psi }=(411\pm 7)~%
\mathrm{MeV}$ are decay constants of the mesons: Their experimental values
are borrowed from Ref.\ \cite{Lakhina:2006vg}.

Besides, one should specify the matrix element $\langle \Upsilon (p^{\prime
},\varepsilon _{1})J/\psi (q,\varepsilon _{2})|T(p,\epsilon )\rangle $ which
can be done by decomposing it in contributions of all possible
Lorentz-invariant terms made of the momenta and polarization tensor and
vectors of the particles $T$, $\Upsilon $ and $J/\psi $ and corresponding
form factors. Then, by requiring the gauge-invariance of the matrix element
it is possible to express $\langle \Upsilon (p^{\prime },\varepsilon
_{1})J/\psi (q,\varepsilon _{2})|T(p,\epsilon )\rangle $ using the
independent form factors (see, for instance, Ref.\ \cite{Aliev:2018kry}). It
turns out that a tensor-vector-vector vertex, in general, contains three
independent form factors which correspond to a pair of vector mesons with
helicities $\lambda =0$, $\pm 1$, and $\pm 2$ \cite%
{Aliev:2018kry,Singer:1983bu,Braun:2000cs,Agaev:2024pil}. In two photon
decays of a tensor meson the main contribution to the width of this process
comes from the amplitude which correspond to a state $\lambda =2$.
Therefore, assuming that the same is true also for the decay $T\rightarrow
J/\psi \Upsilon $, we consider here a pure $\lambda =2$ final state for
which the relevant vertex acquires the following form \cite{Singer:1983bu}:
\begin{eqnarray}
&&\langle \Upsilon (p^{\prime },\varepsilon _{1})J/\psi (q,\varepsilon
_{2})|T(p,\epsilon )\rangle =g_{1}(q^{2})\epsilon _{\tau \rho }^{(\lambda )}
\left[ (\varepsilon _{1}^{\ast }\cdot q)\varepsilon _{2}^{\tau \ast
}p^{\prime \rho }\right.  \notag \\
&&\left. +(\varepsilon _{2}^{\ast }\cdot p^{\prime })\varepsilon _{1}^{\ast
\tau }q^{\rho }-(p^{\prime }\cdot q)\varepsilon _{1}^{\tau \ast }\varepsilon
_{2}^{\rho \ast }-(\varepsilon _{1}^{\ast }\cdot \varepsilon _{2}^{\ast
})p^{\prime \tau }q^{\rho }\right] .  \notag \\
&&  \label{eq:TVV}
\end{eqnarray}%
As a result, for $\Pi _{\mu \nu \alpha \beta }^{\mathrm{Phys}}(p,p^{\prime
}) $ we get the expression
\begin{eqnarray}
&&\Pi _{\mu \nu \alpha \beta }^{\mathrm{Phys}}(p,p^{\prime })=g_{1}(q^{2})%
\frac{\Lambda f_{\Upsilon }m_{\Upsilon }f_{J/\psi }m_{J/\psi }}{\left(
p^{2}-m^{2}\right) (p^{\prime 2}-m_{\Upsilon }^{2})(q^{2}-m_{J/\psi }^{2})}
\notag \\
&&\times \left[ p_{\beta }^{\prime }p_{\alpha }^{\prime }g_{\mu \nu }+\frac{1%
}{2}p_{\mu }p_{\alpha }^{\prime }g_{\beta \nu }+\frac{1}{2m^{2}}p_{\beta
}p_{\nu }p_{\mu }^{\prime }p_{\alpha }^{\prime }\right.  \notag \\
&&\left. +\text{other structures}\right] +\cdots .
\end{eqnarray}

For the QCD side of the sum rule, we obtain
\begin{eqnarray}
&&\Pi _{\mu \nu \alpha \beta }^{\mathrm{OPE}}(p,p^{\prime })=\int
d^{4}xd^{4}ye^{ip^{\prime }y}e^{-ipx}\left\{ \mathrm{Tr}\left[ \gamma _{\mu
}S_{b}^{ia}(y-x)\right. \right.   \notag \\
&&\left. \times \gamma _{\alpha }\widetilde{S}_{c}^{jb}(-x)\gamma _{\nu }%
\widetilde{S}_{c}^{bj}(x)\gamma _{\beta }S_{b}^{ai}(x-y)\right]   \notag \\
&&\left. -\mathrm{Tr}\left[ \gamma _{\mu }S_{b}^{ia}(y-x)\gamma _{\alpha }%
\widetilde{S}_{c}^{jb}(-x)\gamma _{\nu }\widetilde{S}_{c}^{aj}(x)\gamma
_{\beta }S_{b}^{bi}(x-y)\right] \right\} .  \notag \\
&&  \label{eq:CF3}
\end{eqnarray}%
We utilize the structures proportional to $p_{\mu }p_{\alpha }^{\prime
}g_{\beta \nu }$ in the correlators and use corresponding amplitudes $\Pi _{1}^{%
\mathrm{Phys}}(p^{2},p^{\prime 2},q^{2})$ and $\Pi _{1}^{\mathrm{OPE}%
}(p^{2},p^{\prime 2},q^{2})$ to find SR for the form factor $g_{1}(q^{2})$.
After standard operations the sum rule for $g_{1}(q^{2})$ reads
\begin{equation}
g_{1}(q^{2})=\frac{2(q^{2}-m_{J/\psi }^{2})}{\Lambda f_{\Upsilon
}m_{\Upsilon }f_{J/\psi }m_{J/\psi }}e^{m^{2}/M_{1}^{2}}e^{m_{\Upsilon
}^{2}/M_{2}^{2}}\Pi _{1}(\mathbf{M}^{2},\mathbf{s}_{0},q^{2}).
\label{eq:SRG}
\end{equation}%
In Eq.\ (\ref{eq:SRG}), $\Pi _{1}(\mathbf{M}^{2},\mathbf{s}_{0},q^{2})$ is
the Borel transformed and subtracted function $\Pi _{1}^{\mathrm{OPE}%
}(p^{2},p^{\prime 2},q^{2})$. It depends on the parameters $\mathbf{M}%
^{2}=(M_{1}^{2},M_{2}^{2})$ and $\mathbf{s}_{0}=(s_{0},s_{0}^{\prime })$
where the pairs $(M_{1}^{2},s_{0})$ and $(M_{2}^{2},s_{0}^{\prime })$
correspond to the tetraquark and $\Upsilon $ channels, and is given by the
following formula%
\begin{eqnarray}
&&\Pi _{1}(\mathbf{M}^{2},\mathbf{s}_{0},q^{2})=\int_{4\mathcal{M}%
^{2}}^{s_{0}}ds\int_{4m_{b}^{2}}^{s_{0}}ds^{\prime }\rho _{1}(s,s^{\prime
},q^{2})  \notag \\
&&\times e^{-s/M_{1}^{2}-s^{\prime }/M_{2}^{2}}+\mathcal{B}\Pi _{1}^{\mathrm{%
Dim4}}(p^{2},p^{\prime 2},q^{2}).  \label{eq:CorrF1}
\end{eqnarray}%
The explicit expressions of $\rho _{1}(s,s^{\prime },q^{2})$ and $\Pi _{1}^{%
\mathrm{Dim4}}(p^{2},p^{\prime 2},q^{2})$ can be found in the Appendix.

Requirements which should be satisfied by the auxiliary parameters $\mathbf{M%
}^{2}$ and $\mathbf{s}_{0}$ are universal for all SR computations and have
been explained in the previous section. Numerical analysis shows that the
regions in Eq.\ (\ref{eq:Wind1}) for the parameters $(M_{1}^{2},s_{0})$ and
\begin{equation}
M_{2}^{2}\in \lbrack 10,12]~\mathrm{GeV}^{2},\ s_{0}^{\prime }\in \lbrack
98,100]~\mathrm{GeV}^{2}.  \label{eq:Wind3}
\end{equation}%
for $(M_{2}^{2},s_{0}^{\prime })$ satisfy all these requirements. Because
the form factor $g_{1}(q^{2})$ depends on the mass and current coupling of
the tetraquark $T$, this choice for $(M_{1}^{2},s_{0})$ excludes also
additional uncertainties in $m$ and $\Lambda $, as well as in $g_{1}(q^{2})$
which may appear beyond the regions Eq.\ (\ref{eq:Wind1}). It is worth
noting that $s_{0}^{\prime }$ is limited by the mass $m_{\Upsilon
(2S)}=(10023.4\pm 0.5)~\mathrm{MeV}\ $of the radially excited state $%
\Upsilon (2S)$, i.e., $s_{0}^{\prime }<m_{\Upsilon (2S)}^{2}$.

\begin{figure}[h]
\includegraphics[width=8.5cm]{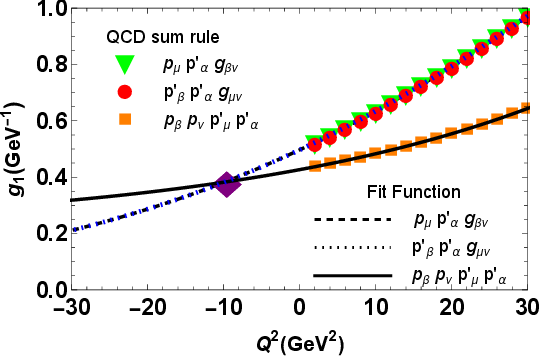}
\caption{Sum rule's data and fit functions for $g_{1}(Q^{2})$. The diamond
fixes the point $Q^{2}=-m_{J/\protect\psi }^{2}$ where $g_{1}$ has been
estimated. }
\label{fig:Fit}
\end{figure}

The SR method leads to reliable predictions for the form factor $%
g_{1}(q^{2}) $ in the Euclidean region $q^{2}<0$. But $g_{1}(q^{2})$
determines the strong coupling $g_{1}$ at the mass shell $q^{2}=m_{J/\psi
}^{2}$. Therefore, it is convenient to introduce the function $g_{1}(Q^{2})$
with $Q^{2}=-q^{2}$ and use it in our analysis. The results obtained for $%
g_{1}(Q^{2})$ are plotted in Fig.\ \ref{fig:Fit}, where $Q^{2}$ varies
inside the limits $Q^{2}=2-30~\mathrm{GeV}^{2}$.

\begin{figure}[h]
\includegraphics[width=8.5cm]{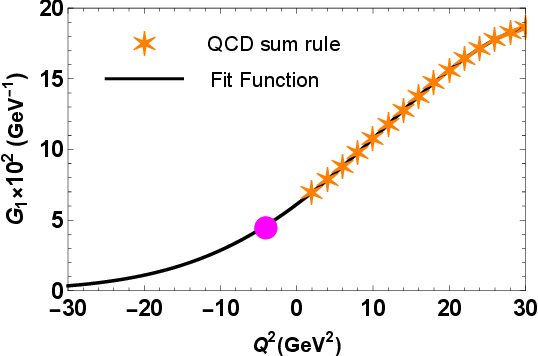}
\caption{Sum rule data and extrapolating function for the form factor $%
G_{1}(Q^{2})$. The circle shows the point $Q^{2}=-m_{D^{\ast 0}}^{2}$. }
\label{fig:Fit1}
\end{figure}

As it has been emphasized above, the strong coupling $g_{1}$ should be
extracted at $q^{2}=m_{J/\psi }^{2}$, i.e., at $Q^{2}=-m_{J/\psi }^{2}$
where the SR method does not work. Therefore, we introduce the fit function $%
\mathcal{G}_{1}(Q^{2})$ that at momenta $Q^{2}>0$ gives the same SR data,
but can be extrapolated to the domain of negative $Q^{2}$. For these
purposes, we utilize the function
\begin{equation}
\mathcal{G}_{i}(Q^{2})=\mathcal{G}_{i}^{0}\mathrm{\exp }\left[ c_{i}^{1}%
\frac{Q^{2}}{m^{2}}+c_{i}^{2}\left( \frac{Q^{2}}{m^{2}}\right) ^{2}\right] ,
\label{eq:FitF}
\end{equation}%
where $\mathcal{G}_{i}^{0}$, $c_{i}^{1}$, and $c_{i}^{2}$ are fitted
constants. Then, having compared QCD output and Eq.\ (\ref{eq:FitF}), it is
easy to find
\begin{equation}
\mathcal{G}_{1}^{0}=0.50~\mathrm{GeV}^{-1},c_{1}^{1}=4.10,\text{and }%
c_{1}^{2}=-2.66.  \label{eq:FF1}
\end{equation}%
This function is also shown in Fig.\ \ref{fig:Fit1}, where a nice agreement
of $\mathcal{G}_{1}(Q^{2})$ and QCD data is clear. For the strong coupling $%
g_{1}$, we find
\begin{equation}
g_{1}\equiv \mathcal{G}_{1}(-m_{J/\psi }^{2})=(3.9\pm 0.9)\times 10^{-1}\
\mathrm{GeV}^{-1}.  \label{eq:g1}
\end{equation}

The form factor $g_{1}(Q^{2})$ and coupling $g_{1}$ can also be extracted
from alternative SRs. To this end, we have used the amplitudes that in the
correlators $\Pi _{\mu \nu \alpha \beta }^{\mathrm{Phys}}(p,p^{\prime })$
and $\Pi _{\mu \nu \alpha \beta }^{\mathrm{OPE}}(p,p^{\prime })$ correspond
to structures $p_{\beta }^{\prime }p_{\alpha }^{\prime }g_{\mu \nu }$ and $%
p_{\beta }p_{\nu }p_{\mu }^{\prime }p_{\alpha }^{\prime }$, respectively.
Numerical predictions for $g_{1}^{\prime }(Q^{2})$ and $g_{1}^{\prime \prime
}(Q^{2})$ found using these new SRs are depicted in Fig.\ \ref{fig:Fit} as
well. It is seen that in the case of the structure $p_{\beta }^{\prime
}p_{\alpha }^{\prime }g_{\mu \nu }$ the sum rule data almost coincide with
ones extracted above for $g_{1}(Q^{2})$. Consequently the parameters of the
extrapolating function $\mathcal{G}_{1}^{\prime }(Q^{2})$, and $%
g_{1}^{\prime }=\mathcal{G}_{1}^{\prime }(-m_{J/\psi }^{2})$ with a high
accuracy are identical to our results from Eqs.\ (\ref{eq:FF1}) and (\ref%
{eq:g1}). The sum rule that corresponds to the structure $p_{\beta }p_{\nu
}p_{\mu }^{\prime }p_{\alpha }^{\prime }$ leads for $g_{1}^{\prime \prime
}(q^{2})$ to different predictions. These SR points can be extrapolated by
employing $\mathcal{G}_{1}^{\prime \prime }(Q^{2})$ with parameters $%
\mathcal{G}_{1}^{0\prime \prime }=0.43~\mathrm{GeV}^{-1},c_{1}^{1\prime
\prime }=1.90,$and $c_{1}^{2\prime \prime }=1.72$. Though QCD data differ
from each other the fitting function $\mathcal{G}_{1}^{\prime \prime
}(Q^{2}) $ gives at the mass shell $Q^{2}=-m_{J/\psi }^{2}$
\begin{equation}
g_{1}^{\prime \prime }\equiv \mathcal{G}_{1}^{\prime \prime }(-m_{J/\psi
}^{2})=(3.8\pm 0.9)\times 10^{-1}\ \mathrm{GeV}^{-1},
\end{equation}%
which is very close to Eq.\ (\ref{eq:g1}). In other words, three different
structures in the correlation functions $\Pi _{\mu \nu \alpha \beta }^{%
\mathrm{Phys}}(p,p^{\prime })$ and $\Pi _{\mu \nu \alpha \beta }^{\mathrm{OPE%
}}(p,p^{\prime })$, and corresponding SRs lead almost to the same result for
the strong coupling $g_{1}$ at the vertex $TJ/\psi \Upsilon $. Because,
uncertainty in $g_{1}$ generated by a choice of the different structures is
considerably smaller than theoretical errors of the SR method itself, it can
be safely neglected.

The partial width of the decay $T\rightarrow J/\psi \Upsilon $ is determined
by the expression%
\begin{equation}
\Gamma \left[ T\rightarrow J/\psi \Upsilon \right] =g_{1}^{2}\frac{\lambda
_{1}}{40\pi m^{2}}|M_{1}|^{2},  \label{eq:PDw2}
\end{equation}%
where%
\begin{eqnarray}
&&|M_{1}|^{2}=\frac{1}{6m^{4}}\left[ m_{J/\psi }^{8}+m_{J/\psi
}^{6}(m^{2}-4m_{\Upsilon }^{2})+(m^{2}-m_{\Upsilon }^{2})^{2}\right.  \notag
\\
&&\times (6m^{4}+3m^{2}m_{\Upsilon }^{2}+m_{\Upsilon }^{4})+m_{J/\psi
}^{4}(m^{4}-m^{2}m_{\Upsilon }^{2}+6m_{\Upsilon }^{4})  \notag \\
&&\left. -m_{J/\psi }^{2}(9m^{6}-34m^{4}m_{\Upsilon }^{2}+m^{2}m_{\Upsilon
}^{4}+4m_{\Upsilon }^{6})\right] ,  \label{eq:M1}
\end{eqnarray}%
and $\lambda _{1}=\lambda (m,m_{\Upsilon },m_{J/\psi })$
\begin{equation}
\lambda (x,y,z)=\frac{\sqrt{%
x^{4}+y^{4}+z^{4}-2(x^{2}y^{2}+x^{2}z^{2}+y^{2}z^{2})}}{2x}.
\end{equation}

Then, we obtain $\ $%
\begin{equation}
\Gamma \left[ T\rightarrow J/\psi \Upsilon \right] =(27.7\pm 9.1)~\mathrm{MeV%
}.  \label{eq:DW2}
\end{equation}

\subsection{Decay $T\rightarrow \protect\eta _{b}\protect\eta _{c}$}


The partial width of the process $T\rightarrow \eta _{b}\eta _{c}$ is
governed by the strong coupling $g_{2}$ at the vertex $T\eta _{b}\eta _{c}$.
In the framework of the SR method the relevant form factor $g_{2}(q^{2})$
can be obtained from analysis of the correlator%
\begin{eqnarray}
\Pi _{\mu \nu }(p,p^{\prime }) &=&i^{2}\int d^{4}xd^{4}ye^{ip^{\prime
}y}e^{-ipx}\langle 0|\mathcal{T}\{\ J^{\eta _{b}}(y)  \notag \\
&&\times J^{\eta _{c}}(0)J_{\mu \nu }^{\dagger }(x)\}|0\rangle .
\label{eq:CF7}
\end{eqnarray}%
The interpolating currents of the quarkonia $\eta _{c}$ and $\eta _{b}$ in
Eq.\ (\ref{eq:CF7}) are
\begin{equation}
J^{\eta _{c}}(x)=\overline{c}_{i}(x)i\gamma _{5}c_{i}(x),\ J^{\eta _{b}}(x)=%
\overline{b}_{j}(x)i\gamma _{5}b_{j}(x).  \label{eq:C3}
\end{equation}

The matrix elements
\begin{eqnarray}
&&\langle 0|J^{\eta _{b}}|\eta _{b}(p^{\prime })\rangle =\frac{f_{\eta
_{b}}m_{\eta _{b}}^{2}}{2m_{b}},  \notag \\
&&\langle 0|J^{\eta _{c}}|\eta _{c}(q)\rangle =\frac{f_{\eta _{c}}m_{\eta
_{c}}^{2}}{2m_{c}}  \label{eq:ME4}
\end{eqnarray}%
are necessary to calculate $\Pi _{\mu \nu }^{\mathrm{Phys}}(p,p^{\prime })$
with $f_{\eta _{b}}$, $m_{\eta _{b}}$ and $f_{\eta _{c}}$, $m_{\eta _{c}}$
being the decay constants and masses of the mesons $\eta _{b}$ and $\eta
_{c} $, respectively. The vertex $T\eta _{b}\eta _{c}$ is given by the
expression \cite{Agaev:2024pil}%
\begin{equation}
\langle \eta _{b}(p^{\prime })\eta _{c}(q)|T(p,\epsilon )\rangle
=g_{2}(q^{2})\epsilon _{\alpha \beta }^{(\lambda )}(p)p^{\prime \alpha
}p^{\prime \beta }.
\end{equation}%
For the correlator $\Pi _{\mu \nu }^{\mathrm{Phys}}(p,p^{\prime })$, we find
\begin{eqnarray}
&&\Pi _{\mu \nu }^{\mathrm{Phys}}(p,p^{\prime })=g_{2}(q^{2})\frac{\Lambda
f_{\eta _{b}}m_{\eta _{b}}^{2}f_{\eta _{c}}m_{\eta _{c}}^{2}}{%
4m_{b}m_{c}\left( p^{2}-m^{2}\right) (p^{\prime 2}-m_{\eta _{b}}^{2})}
\notag \\
&&\times \frac{1}{(q^{2}-m_{\eta _{c}}^{2})}\left[ \frac{m^{4}-2m^{2}(m_{%
\eta _{b}}^{2}+q^{2})+(m_{\eta _{b}}^{2}-q^{2})^{2}}{12m^{2}}g_{\mu \nu
}\right.  \notag \\
&&\left. +p_{\mu }^{\prime }p_{\nu }^{\prime }+\text{other terms}\right] .
\label{eq:CF4}
\end{eqnarray}%
The QCD side of the sum rule $\Pi _{\mu \nu }^{\mathrm{OPE}}(p,p^{\prime })$
has the form
\begin{eqnarray}
&&\Pi _{\mu \nu }^{\mathrm{OPE}}(p,p^{\prime })=i\int
d^{4}xd^{4}ye^{ip^{\prime }y}e^{-ipx}\left\{ \mathrm{Tr}\left[ \gamma
_{5}S_{b}^{ia}(y-x)\right. \right.  \notag \\
&&\left. \times \gamma _{\mu }\widetilde{S}_{c}^{jb}(-x)\gamma _{5}%
\widetilde{S}_{c}^{bj}(x)\gamma _{\nu }S_{b}^{ai}(x-y)\right]  \notag \\
&&\left. -\mathrm{Tr}\left[ \gamma _{5}S_{b}^{ia}(y-x)\gamma _{\mu }%
\widetilde{S}_{c}^{jb}(-x)\gamma _{5}\widetilde{S}_{c}^{aj}(x)\gamma _{\nu
}S_{b}^{bi}(x-y)\right] \right\} .  \notag \\
&&  \label{eq:CF5}
\end{eqnarray}%
The functions $\Pi _{\mu \nu }^{\mathrm{Phys}}(p,p^{\prime })$ and $\Pi
_{\mu \nu }^{\mathrm{OPE}}(p,p^{\prime })$ have the same Lorentz structures.
We consider terms $\sim p_{\mu }^{\prime }p_{\nu }^{\prime }$ and use
corresponding invariant amplitudes $\Pi _{2}^{\mathrm{Phys}}(p^{2},p^{\prime
2},q^{2})$ and $\Pi _{2}^{\mathrm{OPE}}(p^{2},p^{\prime 2},q^{2})$ to derive
the sum rule for the form factor $g_{2}(q^{2})$

\begin{eqnarray}
&&g_{2}(q^{2})=\frac{4m_{c}m_{b}(q^{2}-m_{\eta _{c}}^{2})}{\Lambda f_{\eta
_{c}}m_{\eta _{c}}^{2}f_{\eta _{b}}m_{\eta _{b}}^{2}}%
e^{m^{2}/M_{1}^{2}}e^{m_{\eta _{b}}^{2}/M_{2}^{2}}  \notag \\
&&\times \Pi _{2}(\mathbf{M}^{2},\mathbf{s}_{0},q^{2}),  \label{eq:SRCoup}
\end{eqnarray}%
where $\Pi _{2}(\mathbf{M}^{2},\mathbf{s}_{0},q^{2})$ is the amplitude $\Pi
_{2}^{\mathrm{OPE}}(p^{2},p^{\prime 2},q^{2})$ after Borel transformations
and continuum subtractions.

The remaining manipulations are usual prescriptions of the SR method which
have been explained above. In numerical computations, for the masses of the
quarkonia $\eta _{c}$ and $\eta _{b}$, we use $m_{\eta _{c}}=(2984.1\pm 0.4)~%
\mathrm{MeV}$, $m_{\eta _{b}}=(9398.7~\pm 2.0)\ \mathrm{MeV}\ $ from PDG
\cite{PDG:2022}. The decay constant $f_{\eta _{c}}=(421\pm 35)~\mathrm{MeV}$
was extracted from SR analysis \cite{Veliev:2010vd}, whereas for $f_{\eta
_{b}}$ we employ $724~\mathrm{MeV}$. We have utilized also the following
windows for $M_{2}^{2}$, and $s_{0}^{\prime }$ in the $\eta _{b}$ channel
\begin{equation}
M_{2}^{2}\in \lbrack 10,12]~\mathrm{GeV}^{2},\ s_{0}^{\prime }\in \lbrack
95,99]~\mathrm{GeV}^{2}.
\end{equation}

The extrapolating function $\mathcal{G}_{2}(Q^{2})$ and parameters $\mathcal{%
G}_{2}^{0}=33.63~\mathrm{GeV}^{-1}$, $c_{2}^{1}=8.31$, and $c_{2}^{2}=-13.55$
lead to reasonable agreement with SR data. Then the strong coupling $g_{2}$
amounts to
\begin{equation}
g_{2}\equiv \mathcal{G}_{2}(-m_{\eta _{c}}^{2})=(20.4\pm 4.9)\ \mathrm{GeV}%
^{-1}.
\end{equation}%
The partial width of this process is equal to
\begin{equation}
\Gamma \left[ T\rightarrow \eta _{b}\eta _{c}\right] =g_{2}^{2}\frac{\lambda
_{2}}{40\pi m^{2}}|M_{2}|^{2},
\end{equation}%
where%
\begin{eqnarray}
&&|M_{2}|^{2}=\frac{\left[ m^{4}+(m_{\eta _{b}}^{2}-m_{\eta
_{c}}^{2})^{2}-2m^{2}(m_{\eta _{b}}^{2}+m_{\eta _{c}}^{2})\right] ^{2}}{%
24m^{4}},  \notag \\
&&  \label{eq:M2}
\end{eqnarray}%
and $\lambda _{2}=\lambda (m,m_{\eta _{b}},m_{\eta _{c}})$.

The width of the decay $T\rightarrow \eta _{b}\eta _{c}$ is
\begin{equation}
\Gamma \left[ T\rightarrow \eta _{b}\eta _{c}\right] =(21.1\pm 7.2)~\mathrm{%
MeV}.  \label{eq:DW20}
\end{equation}


\section{Modes $T\rightarrow B_{c}^{\ast +}B_{c}^{\ast -}$ and $%
B_{c}^{+}B_{c}^{-}$}

\label{sec:Widths2}


Here, we consider decays of the tensor tetraquark $T$ to $B_{c}^{\ast
+}B_{c}^{\ast -}$ and $B_{c}^{+}B_{c}^{-}$ final states. It is known that
experimental information about $B_{c}$ mesons is limited by the mass of $%
B_{c}^{\pm }$ and its first radial excitation $B_{c}^{\pm }(2S)$ \cite%
{PDG:2022}. Therefore, for parameters of the $c\overline{b}$ ($b\overline{c}$%
) mesons with other spin-parities one should use theoretical predictions. In
the case of the vector meson $B_{c}^{\ast \pm }$, for its mass and decay
constant we employ
\begin{equation}
m_{B_{c}^{\ast }}=6338~\mathrm{MeV},\ f_{B_{c}^{\ast }}=471~\mathrm{MeV}
\end{equation}%
from Refs.\ \cite{Godfrey:2004ya,Eichten:2019gig}, respectively. We also
utilize the experimental value $m_{B_{c}}=(6274.47\pm 0.27\pm 0.17)~\mathrm{%
MeV}$ for the mass of \ $B_{c}^{\pm }$ and decay constant $f_{B_{c}}=(371\pm
37)~\mathrm{MeV}$ from Ref.\ \cite{Wang:2024fwc}. It is not difficult to see
that the processes $T\rightarrow B_{c}^{\ast +}B_{c}^{\ast -}$ and $%
B_{c}^{+}B_{c}^{-}$ are permitted decay modes of the tensor
diquark-antidiquark state $T$, because thresholds for production of the $%
B_{c}^{\ast +}B_{c}^{\ast -}$ and $B_{c}^{+}B_{c}^{-}$ final-states $12.68~%
\mathrm{GeV}$ and $12.55~\mathrm{GeV}$ are below its mass $m$.


\subsection{$T\rightarrow B_{c}^{\ast +}B_{c}^{\ast -}$}


Analysis of this decay goes in line with a scheme presented and explained
above. Therefore, we write down principal formulas and final results.

The correlation function to derive SR for the form factor $\widetilde{g}%
_{1}(q^{2})$ responsible for strong interaction at the vertex $TB_{c}^{\ast
+}B_{c}^{\ast -}$ is%
\begin{eqnarray}
\widetilde{\Pi }_{\mu \nu \alpha \beta }(p,p^{\prime }) &=&i^{2}\int
d^{4}xd^{4}ye^{ip^{\prime }y}e^{-ipx}\langle 0|\mathcal{T}\{J_{\mu
}^{B_{c}^{\ast +}}(y)  \notag \\
&&\times J_{\nu }^{B_{c}^{\ast -}}(0)J_{\alpha \beta }^{\dagger
}(x)\}|0\rangle .
\end{eqnarray}%
Here, $J_{\mu }^{B_{c}^{\ast +}}$ and $J_{\nu }^{B_{c}^{\ast -}}$ are the
interpolating currents of $B_{c}^{\ast +}$ and $B_{c}^{\ast -}$ mesons which
are determined by the expressions%
\begin{equation}
J_{\mu }^{B_{c}^{\ast +}}(x)=\overline{b}_{i}(x)\gamma _{\mu }c_{i}(x),\
J_{\nu }^{B_{c}^{\ast -}}(x)=\overline{c}_{j}(x)\gamma _{\nu }b_{j}(x).
\end{equation}

In terms of the physical parameters of the particles $\widetilde{\Pi }_{\mu
\nu \alpha \beta }(p,p^{\prime })$ acquires the following form
\begin{eqnarray}
&&\widetilde{\Pi }_{\mu \nu \alpha \beta }(p,p^{\prime })=\frac{\langle
0|J_{\mu }^{B_{c}^{\ast +}}|B_{c}^{\ast +}(p^{\prime },\varepsilon
_{1})\rangle }{p^{\prime 2}-m_{B_{c}^{\ast }}^{2}}\frac{\langle 0|J_{\nu
}^{B_{c}^{\ast -}}|B_{c}^{\ast -}(q,\varepsilon _{2})\rangle }{%
q^{2}-m_{B_{c}^{\ast }}^{2}}  \notag \\
&&\times \langle B_{c}^{\ast +}(p^{\prime },\varepsilon _{1})B_{c}^{\ast
-}(q,\varepsilon _{2})|T(p,\epsilon )\rangle \frac{\langle T(p,\epsilon
)|J_{\alpha \beta }^{\dagger }|0\rangle }{p^{2}-m^{2}}  \notag \\
&&+\cdots .
\end{eqnarray}%
Subsequent calculations are carried out using the matrix elements
\begin{eqnarray}
&&\langle 0|J_{\mu }^{B_{c}^{\ast +}}|B_{c}^{\ast +}(p^{\prime },\varepsilon
_{1})\rangle =f_{B_{c}^{\ast }}m_{B_{c}^{\ast }}\varepsilon _{1\mu
}(p^{\prime }),  \notag \\
&&\langle 0|J_{\nu }^{B_{c}^{\ast -}}|B_{c}^{\ast -}(q,\varepsilon
_{2})\rangle =f_{B_{c}^{\ast }}m_{B_{c}^{\ast }}\varepsilon _{2\nu }(q),
\end{eqnarray}%
where $\varepsilon _{1\mu }(p^{\prime })$ and $\varepsilon _{2\nu }(q)$ are
the polarization vectors of $B_{c}^{\ast +}$ and $B_{c}^{\ast -}$,
respectively. The vertex $TB_{c}^{\ast +}B_{c}^{\ast -}$ is considered in
the form Eq.\ (\ref{eq:TVV}) with replacement $g_{1}(q^{2})\rightarrow
\widetilde{g}_{1}(q^{2})$.

Then $\widetilde{\Pi }_{\mu \nu \alpha \beta }^{\mathrm{Phys}}(p,p^{\prime
}) $ in terms of the physical parameters of the tetraquark $T$ and mesons $%
B_{c}^{\ast \pm }$ reads
\begin{eqnarray}
&&\widetilde{\Pi }_{\mu \nu \alpha \beta }^{\mathrm{Phys}}(p,p^{\prime })=%
\frac{\widetilde{g}_{1}(q^{2})\Lambda f_{B_{c}^{\ast }}^{2}m_{B_{c}^{\ast
}}^{2}}{\left( p^{2}-m^{2}\right) \left( p^{\prime 2}-m_{B_{c}^{\ast
}}^{2}\right) }  \notag \\
&&\times \frac{1}{(q^{2}-m_{B_{c}^{\ast }}^{2})}\left[ \frac{1}{2}p_{\mu
}p_{\alpha }^{\prime }g_{\nu \beta }+\frac{m^{2}+m_{B_{c}^{\ast }}^{2}-q^{2}%
}{4m^{2}}p_{\mu }^{\prime }p_{\alpha }g_{\beta \nu }\right.  \notag \\
&&\left. +p_{\beta }^{\prime }p_{\alpha }^{\prime }g_{\mu \nu }+\text{other
structures}\right] +\cdots .
\end{eqnarray}%
The function $\widetilde{\Pi }_{\mu \nu \alpha \beta }(p,p^{\prime })$
computed in terms of the quark propagators is equal to
\begin{eqnarray}
&&\widetilde{\Pi }_{\mu \nu \alpha \beta }^{\mathrm{OPE}}(p,p^{\prime
})=i\int d^{4}xd^{4}ye^{ip^{\prime }y}e^{-ipx}\left\{ \mathrm{Tr}\left[
\gamma _{\mu }S_{b}^{ia}(y-x)\right. \right.  \notag \\
&&\left. \times \gamma _{\alpha }\widetilde{S}_{c}^{jb}(-x)\gamma _{\nu }%
\widetilde{S}_{b}^{aj}(x)\gamma _{\beta }S_{c}^{bi}(x-y)\right]  \notag \\
&&\left. -\mathrm{Tr}\left[ \gamma _{\mu }S_{b}^{ia}(y-x)\gamma _{\alpha }%
\widetilde{S}_{c}^{jb}(-x)\gamma _{\nu }\widetilde{S}_{b}^{bj}(x)\gamma
_{\beta }S_{c}^{ai}(x-y)\right] \right\} .  \notag \\
&&
\end{eqnarray}

The sum rule for $\widetilde{g}_{1}(q^{2})$
\begin{eqnarray}
&&\widetilde{g}_{1}(q^{2})=\frac{2(q^{2}-m_{\eta _{c}}^{2})}{\Lambda
f_{B_{c}^{\ast }}^{2}m_{B_{c}^{\ast }}^{2}}e^{m^{2}/M_{1}^{2}}e^{m_{B_{c}^{%
\ast }}^{2}/M_{2}^{2}}  \notag \\
&&\times \widetilde{\Pi }_{1}(\mathbf{M}^{2},\mathbf{s}_{0},q^{2})
\end{eqnarray}%
is derived using invariant amplitudes $\widetilde{\Pi }_{1}^{\mathrm{Phys}%
}(p^{2},p^{\prime 2},q^{2})$ and $\widetilde{\Pi }_{1}^{\mathrm{OPE}%
}(p^{2},p^{\prime 2},q^{2})$ which correspond to terms $\sim p_{\mu
}p_{\alpha }^{\prime }g_{\nu \beta }$ in the correlators $\widetilde{\Pi }%
_{\mu \nu \alpha \beta }^{\mathrm{Phys}}(p,p^{\prime })$ and $\widetilde{\Pi
}_{\mu \nu \alpha \beta }^{\mathrm{OPE}}(p,p^{\prime })$, respectively.
Above $\widetilde{\Pi }_{1}(\mathbf{M}^{2},\mathbf{s}_{0},q^{2})$ \ is the
amplitude $\widetilde{\Pi }_{1}^{\mathrm{OPE}}(p^{2},p^{\prime 2},q^{2})$
obtained after relevant transformations.

Numerical computations have been carried out by employing the following
values for the parameters $M_{2}^{2}$ and$\ s_{0}^{\prime }$ in the $%
B_{c}^{\ast +}$ channel
\begin{equation}
M_{2}^{2}\in \lbrack 6.5,7.5]~\mathrm{GeV}^{2},\ s_{0}^{\prime }\in \lbrack
49,51]~\mathrm{GeV}^{2}.
\end{equation}%
The constants of the function $\widetilde{\mathcal{G}}_{1}(Q^{2})$ are equal
to $\widetilde{\mathcal{G}}_{1}^{0}=0.31~\mathrm{GeV}^{-1}$, $\widetilde{c}%
_{1}^{1}=0.19$, and $\widetilde{c}_{1}^{2}=3.37$. We find for the strong
coupling $\widetilde{g}_{1}$
\begin{equation}
\widetilde{g}_{1}\equiv \widetilde{\mathcal{G}}_{1}(-m_{B_{c}^{\ast
}}^{2})=(3.6\pm 0.9)\times 10^{-1}\ \mathrm{GeV}^{-1}.
\end{equation}%
The partial width of the decay $T\rightarrow B_{c}^{\ast +}B_{c}^{\ast -}$
is equal to
\begin{equation}
\Gamma \left[ T\rightarrow B_{c}^{\ast +}B_{c}^{\ast -}\right] =\frac{%
\widetilde{g}_{1}^{2}\widetilde{\lambda }_{1}}{80\pi m^{2}}%
(m^{4}-3m^{2}m_{B_{c}^{\ast }}^{2}+6m_{B_{c}^{\ast }}^{4}),
\end{equation}%
where $\widetilde{\lambda }_{1}=\lambda (m,m_{B_{c}^{\ast }},m_{B_{c}^{\ast
}})$. Alternatively, the width of this decay can be obtained from Eqs.\ (\ref%
{eq:PDw2}) and (\ref{eq:M1}) upon replacement $m_{J/\psi }=m_{\Upsilon
}\rightarrow m_{B_{c}^{\ast }}$.

Numerical calculations yield
\begin{equation}
\Gamma \left[ T\rightarrow B_{c}^{\ast +}B_{c}^{\ast -}\right] =(20.1\pm
6.6)~\mathrm{MeV}.
\end{equation}


\subsection{$T\rightarrow B_{c}^{+}B_{c}^{-}$}


The process $T\rightarrow B_{c}^{+}B_{c}^{-}$ is investigated in analogous
manner. We consider the correlation function
\begin{eqnarray}
\widetilde{\Pi }_{\mu \nu }(p,p^{\prime }) &=&i^{2}\int
d^{4}xd^{4}ye^{ip^{\prime }y}e^{-ipx}\langle 0|\mathcal{T}\{\
J^{B_{c}^{+}}(y)  \notag \\
&&\times J^{B_{c}^{-}}(0)J_{\mu \nu }^{\dagger }(x)\}|0\rangle ,
\end{eqnarray}%
with $\ J^{B_{c}^{+}}(x)$ and $J^{B_{c}^{-}}(x)$ being the interpolating
currents of the $B_{c}^{+}$ and $B_{c}^{-}$ mesons
\begin{equation}
\ J^{B_{c}^{+}}(x)=\overline{b}_{i}(x)i\gamma _{5}c_{i}(x),\
J^{B_{c}^{-}}(x)=\overline{c}_{j}(x)i\gamma _{5}b_{j}(x).
\end{equation}

The matrix elements of the $B_{c}^{\pm }$ mesons are
\begin{equation}
\langle 0|J^{B_{c}^{\pm }}|B_{c}^{\pm }\rangle =\frac{f_{B_{c}}m_{B_{c}}^{2}%
}{m_{b}+m_{c}}.
\end{equation}%
The vertex $\langle B_{c}^{+}(p^{\prime })B_{c}^{-}(q)|T(p,\epsilon )\rangle
$ has the form
\begin{equation}
\langle B_{c}^{+}(p^{\prime })B_{c}^{-}(q)|T(p,\epsilon )\rangle =\widetilde{%
g}_{2}(q^{2})\epsilon _{\alpha \beta }^{(\lambda )}(p)p^{\prime \alpha
}p^{\prime \beta }.
\end{equation}%
The $\widetilde{\Pi }_{\mu \nu }^{\mathrm{Phys}}(p,p^{\prime })$ obtained
using these matrix elements after some substitutions [$m_{\eta _{b}}^{2} %
\mbox{and } m_{\eta _{c}}^{2}\rightarrow m_{B_{c}}^{2}$, $%
4m_{c}m_{b}\rightarrow (m_{b}+m_{c})^{2}$, etc.] is given by Eq.\ (\ref%
{eq:CF4}), whereas the QCD side of SR is defined by Eq.\ (\ref{eq:CF5}). The
SR for the form factor $\widetilde{g}_{2}(q^{2})$ is
\begin{eqnarray}
&&\widetilde{g}_{2}(q^{2})=\frac{(m_{b}+m_{c})^{2}(q^{2}-m_{B_{c}}^{2})}{%
\Lambda f_{B_{c}}^{2}m_{B_{c}}^{4}}%
e^{m^{2}/M_{1}^{2}}e^{m_{B_{c}}^{2}/M_{2}^{2}}  \notag \\
&&\times \widetilde{\Pi }_{2}(\mathbf{M}^{2},\mathbf{s}_{0},q^{2}).
\end{eqnarray}

In numerical analysis, we have used the following parameters
\begin{equation}
M_{2}^{2}\in \lbrack 6.5,7.5]~\mathrm{GeV}^{2},\ s_{0}^{\prime }\in \lbrack
45,47]~\mathrm{GeV}^{2}.
\end{equation}%
Computations of the form factor $\widetilde{g}_{2}(q^{2})$ and coupling $%
\widetilde{g}_{2}$ lead to the prediction
\begin{equation}
\widetilde{g}_{2}\equiv \widetilde{\mathcal{G}}_{2}(-m_{B_{c}}^{2})=(26.6\pm
6.4)\ \mathrm{GeV}^{-1},
\end{equation}%
where $\widetilde{\mathcal{G}}_{2}(Q^{2})$ is the fitting function with
parameters $\widetilde{\mathcal{G}}_{2}^{0}=29.89~\mathrm{GeV}^{-1}$, $%
\widetilde{c}_{2}^{1}=0.95$, and $\widetilde{c}_{2}^{2}=1.90$.

The width of the decay $T\rightarrow B_{c}^{+}B_{c}^{-}$ can be computed by
means of the expression
\begin{equation}
\Gamma \left[ T\rightarrow B_{c}^{+}B_{c}^{-}\right] =\widetilde{g}_{2}^{2}%
\frac{\widetilde{\lambda }_{2}}{960\pi m^{2}}\left(
m^{2}-4m_{B_{c}}^{2}\right) ^{2},  \label{eq:PDw3}
\end{equation}%
where $\widetilde{\lambda }_{2}=\lambda (m,m_{B_{c}},m_{B_{c}})$. This
formula in the limit $m_{\eta _{b}} \mbox{and } m_{\eta _{c}}\rightarrow
m_{B_{c}}$ can be obtained from Eq.\ (\ref{eq:M2}). Our computations yield
\begin{equation}
\Gamma \left[ T\rightarrow B_{c}^{+}B_{c}^{-}\right] =(20.1\pm 6.9)~\mathrm{%
MeV}.
\end{equation}


\section{Decays due to $b\overline{b}$ annihilations}

\label{sec:Widths3}


It has been emphasized above that the tetraquark $T$ can transform to
conventional mesons also due to annihilation of $b\overline{b}$ to light
quark-antiquark pairs \cite{Becchi:2020mjz,Becchi:2020uvq,Agaev:2023ara} and
creation of $DD$ mesons with required electric charges and spin-parities.
Here, we consider decays of the tetraquark $T$ to $D^{\ast 0}\overline{D}%
^{\ast 0}$, $D^{0}\overline{D}^{0}$, $D^{\ast +}D^{\ast -}$, and $D^{+}D^{-}$
mesons.

It is clear that these decays are kinematically possible modes for
transformation of the tetraquark $T$ to ordinary mesons. We study these
processes in the same context of the three-point sum rule approach. But here
we encounter a situation when relevant correlation functions contain$\
\overline{b}b$ quarks' vacuum matrix element $\langle \overline{b}b\rangle $
\cite{Agaev:2023ara}. In calculations, we replace this matrix element with
known value of the gluon condensate $\langle \alpha _{s}G^{2}/\pi \rangle $.


\subsection{Decays $T\rightarrow D^{\ast 0}\overline{D}^{\ast 0}$ and $D^{0}%
\overline{D}^{0}$}


Let us analyze the process $T\rightarrow D^{\ast 0}\overline{D}^{\ast 0}$.
To find the coupling $G_{1}$ of particles at the vertex $TD^{\ast 0}%
\overline{D}^{\ast 0}$, we start from the correlation function%
\begin{eqnarray}
\widehat{\Pi }_{\mu \nu \alpha \beta }(p,p^{\prime }) &=&i^{2}\int
d^{4}xd^{4}ye^{ip^{\prime }y}e^{-ipx}\langle 0|\mathcal{T}\{J_{\mu }^{%
\overline{D}^{\ast 0}}(y)  \notag \\
&&\times J_{\nu }^{D^{\ast 0}}(0)J_{\alpha \beta }^{\dagger }(x)\}|0\rangle ,
\label{eq:CF1A}
\end{eqnarray}%
where $J_{\mu }^{\overline{D}^{\ast 0}}(x)$ and $J_{\nu }^{D^{\ast 0}}(x)$
are interpolating currents for the mesons $\overline{D}^{\ast 0}$ and $%
D^{\ast 0}$
\begin{equation}
J_{\mu }^{\overline{D}^{\ast 0}}(x)=\overline{c}_{i}(x)\gamma _{\mu
}u_{i}(x),\text{ }J_{\nu }^{D^{\ast 0}}(x)=\overline{u}_{j}(x)\gamma _{\nu
}c_{j}(x).  \label{eq:CRB}
\end{equation}

The expression of the correlation function $\widehat{\Pi }_{\mu \nu \alpha
\beta }(p,p^{\prime })$ in terms of $T$, $D^{\ast 0}$, and $\overline{D}%
^{\ast 0}$ particles' parameters reads
\begin{eqnarray}
&&\widehat{\Pi }_{\mu \nu \alpha \beta }^{\mathrm{Phys}}(p,p^{\prime })=%
\frac{\langle 0|J_{\mu }^{\overline{D}^{\ast 0}}|\overline{D}^{\ast
0}(p^{\prime },\varepsilon _{1})\rangle }{p^{\prime 2}-m_{D^{\ast 0}}^{2}}%
\frac{\langle 0|J_{\nu }^{D^{\ast 0}}|D^{\ast 0}(q,\varepsilon _{2})\rangle
}{q^{2}-m_{D^{\ast 0}}^{2}}  \notag  \label{eq:CF2} \\
&&\times \langle \overline{D}^{\ast 0}(p^{\prime },\varepsilon _{1})D^{\ast
0}(q,\varepsilon _{2})|T(p,\epsilon )\rangle \frac{\langle T(p,\epsilon
)|J_{\mu }^{\dagger }|0\rangle }{p^{2}-m^{2}}+\cdots ,  \notag \\
&&
\end{eqnarray}%
where $m_{D^{\ast 0}}=(2006.85\pm 0.05)~\mathrm{MeV}$ is the mass of the
mesons $\overline{D}^{\ast 0}$ and $D^{\ast 0}$, whereas $\varepsilon _{1\mu
}$ and $\varepsilon _{2\nu }$ are their polarization vectors.

The matrix elements which are required to calculate $\widehat{\Pi }_{\mu \nu
\alpha \beta }^{\mathrm{Phys}}(p,p^{\prime })$ are
\begin{eqnarray}
\langle 0|J_{\mu }^{\overline{D}^{\ast 0}}|\overline{D}^{\ast 0}(p^{\prime
},\varepsilon _{1})\rangle &=&f_{D^{\ast }}m_{D^{\ast 0}}\varepsilon _{1\mu
}(p^{\prime }),  \notag \\
\langle 0|J_{\nu }^{D^{\ast 0}}|D^{\ast 0}(q,\varepsilon _{2})\rangle
&=&f_{D^{\ast }}m_{D^{\ast 0}}\varepsilon _{2\nu }(q),  \label{eq:ME2B}
\end{eqnarray}%
with $f_{D^{\ast }}=(252.2\pm 22.66)~\mathrm{MeV}$ being the decay constant
of the mesons $D^{\ast 0}$ and $\overline{D}_{D}^{\ast 0}$. The vertex $%
\langle \overline{D}^{\ast 0}(p^{\prime },\varepsilon _{1})D^{\ast
0}(q,\varepsilon _{2})|T(p,\epsilon )\rangle $ is modeled in the form of
Eq.\ (\ref{eq:TVV}).

The correlator $\widehat{\Pi }_{\mu \nu \alpha \beta }^{\mathrm{Phys}%
}(p,p^{\prime })$ is a sum of different components. The SR for the form
factor $G_{1}(q^{2})$ is obtained by employing the invariant amplitude $%
\widehat{\Pi }_{1}^{\mathrm{Phys}}(p^{2},p^{\prime 2},q^{2})$ that
corresponds to the structure $p_{\mu }p_{\alpha }^{\prime }g_{\beta \nu }$.
The same correlation function $\widehat{\Pi }_{\mu \nu \alpha \beta
}(p,p^{\prime })$ computed using the heavy and light quark propagators is
\begin{eqnarray}
&&\widehat{\Pi }_{\mu \nu \alpha \beta }^{\mathrm{OPE}}(p,p^{\prime })=\frac{%
2}{3}\langle \overline{b}b\rangle \int d^{4}xd^{4}ye^{ip^{\prime }y}e^{-ipx}
\notag \\
&&\times \mathrm{Tr}\left[ \gamma _{\mu }S_{u}^{ij}(y)\gamma _{\nu
}{}S_{c}^{jb}(-x)\gamma _{\alpha }\gamma _{\beta }S_{c}^{bi}(x-y)\right] ,
\label{eq:QCDsideA}
\end{eqnarray}%
where $S_{u}(x)$ is the $u$ quark's propagator \cite{Agaev:2020zad}. In what
follows, the function $\widehat{\Pi }_{1}^{\mathrm{OPE}}(p^{2},p^{\prime
2},q^{2})$ is the invariant amplitude that corresponds in $\widehat{\Pi }%
_{\mu \nu \alpha \beta }^{\mathrm{OPE}}(p,p^{\prime })$ to the term $p_{\mu
}p_{\alpha }^{\prime }g_{\beta \nu }$.

For further studies, we make use of the relation between condensates
\begin{equation}
\langle \overline{b}b\rangle =-\frac{1}{12m_{b}}\langle \frac{\alpha
_{s}G^{2}}{\pi }\rangle  \label{eq:Conden}
\end{equation}%
derived in Ref.\ \cite{Shifman:1978bx} from the sum rule analysis. This
expression was obtained at the leading order of the perturbative QCD and is
valid as far as higher order corrections in $m_{b}^{-1}$ are very small.

The SR for the coupling $G_{1}(q^{2})$ reads%
\begin{eqnarray}
&&G_{1}(q^{2})=\frac{2(q^{2}-m_{D^{\ast 0}}^{2})}{\Lambda f_{D^{\ast
}}^{2}m_{D^{\ast 0}}^{2}}e^{m^{2}/M_{1}^{2}}e^{m_{D^{\ast 0}}^{2}/M_{2}^{2}}
\notag \\
&&\times \widehat{\Pi }_{1}(\mathbf{M}^{2},\mathbf{s}_{0},q^{2}),
\end{eqnarray}%
where $\widehat{\Pi }_{1}(\mathbf{M}^{2},\mathbf{s}_{0},q^{2})$ is the
amplitude $\widehat{\Pi }_{1}^{\mathrm{OPE}}(p^{2},p^{\prime 2},q^{2})$
undergone to Borel transformations and continuum subtractions.

To extract $G_{1}(q^{2})$ from this SR, we carry out standard manipulations,
and skip further details: In the $\overline{D}^{\ast 0}$ meson channel, we
have used the parameters
\begin{equation}
M_{2}^{2}\in \lbrack 2,3]~\mathrm{GeV}^{2},\ s_{0}^{\prime }\in \lbrack
5.7,5.8]~\mathrm{GeV}^{2}.  \label{eq:Wind2}
\end{equation}%
The coupling $G_{1}$ has been evaluated by employing SR data for $%
Q^{2}=2-30\ \mathrm{GeV}^{2}$ and the extrapolating function with parameters
$\widehat{\mathcal{G}}_{1}^{0}=0.06~\mathrm{GeV}^{-1}$, $\widehat{c}%
_{1}^{1}=10.67$, and $\widehat{c}_{1}^{2}=-25.14$. The SR data and fit
function $\widehat{\mathcal{G}}_{1}(Q^{2})$ are plotted in Fig.\ \ref%
{fig:Fit1}. The coupling $G_{1}$ has been computed at the mass shell $%
q^{2}=m_{D^{\ast 0}}^{2}$ and amounts to
\begin{equation}
G_{1}\equiv \widehat{\mathcal{G}}_{1}(-m_{D^{\ast 0}}^{2})=(4.62\pm
1.11)\times 10^{-2}\ \mathrm{GeV}^{-1}.  \label{eq:G1}
\end{equation}%
The width of the decay $T\rightarrow D^{\ast 0}\overline{D}^{\ast 0}$ is
\begin{equation}
\Gamma \left[ T\rightarrow D^{\ast 0}\overline{D}^{\ast 0}\right] =(7.7\pm
2.6)~\mathrm{MeV}.
\end{equation}

The second process $T\rightarrow D^{0}\overline{D}^{0}$ is considered
starting from the correlator
\begin{eqnarray}
\widehat{\Pi }_{\mu \nu }(p,p^{\prime }) &=&i^{2}\int
d^{4}xd^{4}ye^{ip^{\prime }y}e^{-ipx}\langle 0|\mathcal{T}\{J^{\overline{D}%
^{0}}(y)  \notag \\
&&\times J^{D^{0}}(0)J_{\mu \nu }^{\dagger }(x)\}|0\rangle ,
\end{eqnarray}%
where the currents $J^{\overline{D}^{0}}(x)$ and $J^{D^{0}}(x)$ are defined
by expressions%
\begin{equation}
J^{\overline{D}^{0}}(x)=\overline{c}_{i}(x)i\gamma _{5}u_{i}(x),\text{ }%
J^{D^{0}}(x)=\overline{u}_{j}(x)i\gamma _{5}c_{j}(x).
\end{equation}%
To get the sum rule for the form factor $G_{2}(q^{2})$ responsible for
strong interaction of particles at the vertex $TD^{0}\overline{D}^{0}$, we
calculate $\widehat{\Pi }_{\mu \nu }^{\mathrm{Phys}}(p,p^{\prime })$ and $%
\widehat{\Pi }_{\mu \nu }^{\mathrm{OPE}}(p,p^{\prime })$.

We determine $\widehat{\Pi }_{\mu \nu }^{\mathrm{Phys}}(p,p^{\prime })$
using the following matrix elements
\begin{equation}
\langle 0|J^{\overline{D}^{0}}|\overline{D}^{0}\rangle =\langle
0|J^{D^{0}}|D^{0}\rangle =\frac{f_{D}m_{D^{0}}^{2}}{m_{c}},
\end{equation}%
and
\begin{equation}
\langle \overline{D}^{0}(p^{\prime })D^{0}(q)|T(p,\epsilon )\rangle
=G_{2}(q^{2})\epsilon _{\alpha \beta }^{(\lambda )}(p)p^{\prime \alpha
}p^{\prime \beta },
\end{equation}%
with $m_{D^{0}}=(1864.84\pm 0.05)~\mathrm{MeV}$ and $f_{D}=(211.9\pm 1.1)~%
\mathrm{MeV}$ being the mass and decay constant of mesons $D^{0}$ and $%
\overline{D}^{0}$ \cite{PDG:2022,Rosner:2015wva}. As a result, we obtain
\begin{eqnarray}
&&\widehat{\Pi }_{\mu \nu }^{\mathrm{Phys}}(p,p^{\prime })=\frac{%
G_{2}(q^{2})\Lambda f_{D}^{2}m_{D^{0}}^{4}}{m_{c}^{2}\left(
p^{2}-m^{2}\right) \left( p^{\prime 2}-m_{D^{0}}^{2}\right) \left(
q^{2}-m_{D^{0}}^{2}\right) }  \notag \\
&&\times \left[ \frac{%
m^{4}-2m^{2}(m_{D^{0}}^{2}+q^{2})+(m_{D^{0}}^{2}-q^{2})^{2}}{12m^{2}}g_{\mu
\nu }\right.  \notag \\
&&\left. +p_{\mu }^{\prime }p_{\nu }^{\prime }-\frac{%
m^{2}+m_{D^{0}}^{2}-q^{2}}{2m^{2}}p_{\mu }p_{\nu }^{\prime }+\text{other
terms}\right] .
\end{eqnarray}%
For $\widehat{\Pi }_{\mu \nu }^{\mathrm{OPE}}(p,p^{\prime })$, we find%
\begin{eqnarray}
&&\widehat{\Pi }_{\mu \nu }^{\mathrm{OPE}}(p,p^{\prime })=\frac{2}{3}\langle
\overline{b}b\rangle \int d^{4}xd^{4}ye^{ip^{\prime }y}e^{-ipx}  \notag \\
&&\times \mathrm{Tr}\left[ \gamma _{5}S_{u}^{ij}(y){}\gamma
_{5}S_{c}^{jb}(-x)\gamma _{\mu }\gamma _{\nu }S_{c}^{bi}(x-y)\right] .
\end{eqnarray}%
We extract SR for $G_{2}(q^{2})$ using the amplitudes $\widehat{\Pi }_{2}^{%
\mathrm{Phys}}(p^{2},p^{\prime 2},q^{2})$ and $\widehat{\Pi }_{2}^{\mathrm{%
OPE}}(p^{2},p^{\prime 2},q^{2})$ corresponding to structures $p_{\mu
}^{\prime }p_{\nu }^{\prime }$ and get
\begin{eqnarray}
&&G_{2}(q^{2})=\frac{m_{c}^{2}(q^{2}-m_{D^{0}}^{2})}{\Lambda
f_{D}^{2}m_{D^{0}}^{4}}e^{m^{2}/M_{1}^{2}}e^{m_{D^{0}}^{2}/M_{2}^{2}}  \notag
\\
&&\times \widehat{\Pi }_{2}(\mathbf{M}^{2},\mathbf{s}_{0},q^{2}),
\end{eqnarray}%
with $\widehat{\Pi }_{2}(\mathbf{M}^{2},\mathbf{s}_{0},q^{2})$ being the
transformed function $\widehat{\Pi }_{2}^{\mathrm{OPE}}(p^{2},p^{\prime
2},q^{2})$.

In numerical calculations we employed the parameters
\begin{equation}
M_{2}^{2}\in \lbrack 1.5,3]~\mathrm{GeV}^{2},\ s_{0}^{\prime }\in \lbrack
5,5.2]~\mathrm{GeV}^{2}.
\end{equation}%
We have found the coupling $G_{2}$ by means of the of the function $\widehat{%
\mathcal{G}}_{2}(Q^{2})$ with $\widehat{\mathcal{G}}_{2}^{0}=0.20~\mathrm{GeV%
}^{-1}$, $\widehat{c}_{2}^{1}=10.72$, and $\widehat{c}_{2}^{2}=-26.80$
\begin{equation}
G_{2}\equiv \widehat{\mathcal{G}}_{2}(-m_{D^{0}}^{2})=(0.16\pm 0.04)\
\mathrm{GeV}^{-1}.
\end{equation}%
The partial width of the decay $T\rightarrow D^{0}\overline{D}^{0}$ is equal
to
\begin{equation}
\Gamma \left[ T\rightarrow D^{0}\overline{D}^{0}\right] =(6.5\pm 2.3)~%
\mathrm{MeV}.
\end{equation}


\subsection{Processes $T\rightarrow D^{\ast +}D^{\ast -}$ and $D^{+}D^{-}$}


The modes $T\rightarrow D^{\ast +}D^{\ast -}$ and $D^{+}D^{-}$ are explored
in accordance with the scheme explained above. Let us analyze the process $%
T\rightarrow D^{\ast +}D^{\ast -}$. The strong form factor $G_{3}(q^{2})$ at
the vertex $TD^{\ast +}D^{\ast -}$ is extracted from the correlation
function
\begin{eqnarray}
\Pi _{\mu \nu \alpha \beta }^{\prime }(p,p^{\prime }) &=&i^{2}\int
d^{4}xd^{4}ye^{ip^{\prime }y}e^{-ipx}\langle 0|\mathcal{T}\{J_{\mu
}^{D^{\ast +}}(y)  \notag \\
&&\times J_{\nu }^{D^{\ast -}}(0)J_{\alpha \beta }^{\dagger }(x)\}|0\rangle ,
\end{eqnarray}%
where currents for the mesons $D^{\ast +}$ and $D^{\ast -}$ are given by the
formulas
\begin{equation}
J_{\mu }^{D^{\ast +}}(x)=\overline{d}_{i}(x)\gamma _{\mu }c_{i}(x),\ J_{\nu
}^{D^{\ast -}}(x)=\overline{d}_{j}(x)\gamma _{\nu }d_{j}(x).
\end{equation}%
The matrix elements of these particles and the vertex are similar to ones
introduced above. Therefore, we omit these expressions and write down the
QCD side of the SR
\begin{eqnarray}
&&\overline{\Pi }_{\mu \nu \alpha \beta }^{\prime \mathrm{OPE}}(p,p^{\prime
})=-\frac{i}{18m_{b}}\langle \frac{\alpha _{s}G^{2}}{\pi }\rangle \int
d^{4}xd^{4}ye^{ip^{\prime }y}e^{-ipx}  \notag \\
&&\times \mathrm{Tr}\left[ \gamma _{\mu }S_{d}^{ij}(y)\gamma _{\nu
}{}S_{c}^{jb}(-x)\gamma _{\alpha }\gamma _{\beta }S_{c}^{bj}(x-y)\right] .
\end{eqnarray}

As usual, we utilize invariant amplitudes corresponding to the structures $%
p_{\mu }p_{\alpha }^{\prime }g_{\beta \nu }$. In numerical calculations the
Borel and continuum subtraction parameters in the $D^{\ast +}$ channel are
fixed as in Eq.\ (\ref{eq:Wind2}). The mass of $D^{\ast \pm }$ mesons is $%
m_{D^{\ast }}=(2010.26\pm 0.05)~\mathrm{MeV}$, whereas for their decay
constants we use $f_{D^{\ast }}=(252.2\pm 22.66)~\mathrm{MeV}$.

The function $\widehat{\mathcal{G}}_{3}(Q^{2})$ with the constants $\widehat{%
\mathcal{G}}_{3}^{0}=0.06~\mathrm{GeV}^{-1}$, $\widehat{c}_{3}^{1}=10.65$,
and $\widehat{c}_{3}^{2}=-25.11$ leads to coupling $G_{3}$
\begin{equation}
G_{3}\equiv \widehat{\mathcal{G}}_{3}(-m_{D^{\ast }}^{2})=(4.63\pm
1.11)\times 10^{-2}\ \mathrm{GeV}^{-1}.
\end{equation}%
For the partial width of the mode $T\rightarrow D^{\ast +}D^{\ast -}$, we
get
\begin{equation}
\Gamma \left[ T\rightarrow D^{\ast +}D^{\ast -}\right] =(7.7\pm 2.7)~\mathrm{%
MeV}.
\end{equation}

The process $T\rightarrow D^{+}D^{-}$ is explored in similar way. The
coupling $G_{4}$ describing the strong interaction of the particles at the
vertex $TD^{+}D^{-}$ is%
\begin{equation}
G_{4}\equiv \widetilde{\mathcal{G}}_{4}(-m_{D}^{2})=(0.16\pm 0.04)~\mathrm{%
GeV}^{-1}.
\end{equation}%
For the width of this decay, we find
\begin{equation}
\Gamma \left[ T\rightarrow D^{+}D^{-}\right] =(6.5\pm 2.3)~\mathrm{MeV}.
\end{equation}

Computations performed in present paper allow us to estimate the full width
of the axial-vector tetraquark $T$ with content $bc\overline{b}\overline{c}$%
. As a result, we obtain%
\begin{equation}
\Gamma \left[ T\right] =(117.4\pm 15.9)~\mathrm{MeV}.
\end{equation}


\section{Conclusions}

\label{sec:Conc}


In present article, we have calculated the mass and full width of the tensor
tetraquark $bc\overline{b}\overline{c}$. Analyses have been performed in the
framework of QCD sum rule method. To evaluate the mass of $T$, we have
applied the two-point SR method, whereas its decays have been studied by
invoking the three-point SR approach.

The mass $m$ of the tensor tetraquark $T$ was evaluated in different
articles, sometimes with contradictory results \cite%
{Faustov:2022mvs,Wu:2016vtq,Liu:2019zuc,Chen:2019vrj,Bedolla:2019zwg,Cordillo:2020sgc,Weng:2020jao,Yang:2021zrc}%
. Our prediction $m=(12.70\pm 0.09)~\mathrm{GeV}$ is smaller than those
reported in publications \cite%
{Faustov:2022mvs,Wu:2016vtq,Liu:2019zuc,Chen:2019vrj}. In Refs. \cite%
{Bedolla:2019zwg,Cordillo:2020sgc,Weng:2020jao,Yang:2021zrc} the authors
found the mass of this state in most of cases below $m$. Thus, $m$ evaluated
in the present work is somewhere between these two groups of predictions.

The results of current paper demonstrate that the tensor state $T$ can decay
to ordinary mesons through the strong fall-apart mechanism. In almost all
articles cited above authors made similar conclusions: Only in Ref. \cite%
{Yang:2021zrc} $T$ was predicted to be stable against two-meson strong
dissociations. But let us emphasize that structures $bc\overline{b}\overline{%
c}$, due to $b\overline{b}$ and $c\overline{c}$ annihilations and
generations of ordinary heavy-light mesons, are always strong-interaction
unstable particles.

We have calculated partial widths of the four processes $T\rightarrow J/\psi
\Upsilon $, $\eta _{b}\eta _{c}$ and $B_{c}^{(\ast )+}B_{c}^{(\ast )-}$
which are dominant decay channels of $T$. We have evaluated also widths of
modes triggered by $b\overline{b}$ annihilations inside of $T$ and
containing at the final states $D^{(\ast )+}D^{(\ast )-}$ and $D^{(\ast )0}%
\overline{D}^{(\ast )0}$ mesons. It is worth noting that the contribution of
these processes is not small and forms approximately $24\%$ of the $T$
tetraquark's full width.

Our predictions characterize $T$ as a wide diquark-antidiquark state, which
can decay to two-meson final states through both fall-apart and $b\overline{b%
}$ annihilation mechanisms. The tensor tetraquark $T$, as well as the scalar
and axial-vector tetraquarks $bc\overline{b}\overline{c}$ establish a family
of fully heavy exotic mesons with different spin-parities. Having compared $%
m $ with masses of the scalar and axial-vector states $m_{\mathrm{S}%
}=(12.697\pm 0.090)~\mathrm{GeV}$ and $m_{\mathrm{AV}}=(12.715\pm 0.090)~%
\mathrm{GeV}$ \cite{Agaev:2024wvp,Agaev:2024mng}, one sees that they form
almost a degenerate system of particles.

Heavy tetraquarks are an inseparable part of the exotic hadron spectroscopy.
Structures $bc\overline{b}\overline{c}$ were not observed yet, but they can
be seen in the future runs of the LHC and Future Circular Collider \cite%
{Carvalho:2015nqf,Abreu:2023wwg}. Publications devoted to fully heavy
four-quark states are concentrated on analysis of their masses. Decays of
these states, including $bc\overline{b}\overline{c}$ ones, did not become
objects of detailed investigations. But besides masses, all conclusions
about nature of discovered resonances have to be also based on knowledge
about their decay channels and widths: This information is required for
reliable interpretation of collected data and for planning new measurements.

\begin{widetext}
\appendix*

\section{ The correlation functions $\Pi (M^{2},s_{0})$ and $\Pi _{1}(%
\mathbf{M}^{2},\mathbf{s}_{0},q^{2})$}

\renewcommand{\theequation}{\Alph{section}.\arabic{equation}} \label{sec:App}


This Appendix contains the correlation functions $\Pi (M^{2},s_{0})$ and $%
\Pi _{1}(\mathbf{M}^{2},\mathbf{s}_{0},q^{2})$, which have been applied to
compute the mass of the tensor tetraquark $T$ and partial width of the decay
$T\rightarrow J/\psi \Upsilon $.

The correlation function $\Pi (M^{2},s_{0})$ which appears in the
SRs has been presented in Eq.\ (\ref{eq:CorrF}):
\begin{equation*}
\Pi (M^{2},s_{0})=\int_{4\mathcal{M}^{2}}^{s_{0}}ds\rho ^{\mathrm{OPE}%
}(s)e^{-s/M^{2}}+\Pi (M^{2}).
\end{equation*}

The components of the spectral density $\rho ^{\mathrm{OPE}}(s)$ are given
by the general expression%
\begin{equation}
\rho (s)=\int_{0}^{1}d\alpha \int_{0}^{1-\alpha }d\beta \int_{0}^{1-\alpha
-\beta }d\gamma \rho (s,\alpha ,\beta ,\gamma ),  \label{eq:A4}
\end{equation}%
where $\alpha $, $\beta $, and $\gamma $ are the Feynman parameters. The
function $\Pi (M^{2})$ is also determined by an Eq.\ (\ref{eq:A4})-type formula
with the integrand $\Pi (M^{2},\alpha ,\beta ,\gamma )$.

The perturbative function $\rho ^{\mathrm{pert.}}(s,\alpha ,\beta ,\gamma )$
is given by the formula
\begin{eqnarray}
&&\rho ^{\mathrm{pert.}}(s,\alpha ,\beta ,\gamma )=\frac{N^{2}\theta ({N)}}{%
512C^{4}A^{4}\pi ^{6}}\left\{ 12C^{2}L_{1}^{3}s^{2}\alpha ^{3}\beta
^{3}\gamma ^{3}+4A^{2}Cs\alpha \beta \gamma (3BCm_{b}m_{c}-4L_{1}^{2}N\alpha
\beta \gamma )\right.  \notag \\
&&\left. +A^{4}(-4Cm_{b}m_{c}(3Cm_{b}m_{c}+N\alpha \beta )+N\gamma
L_{1}(4Cm_{b}m_{c}+N\alpha \beta ))\right\} .
\end{eqnarray}
Here%
\begin{equation}
N=-C\left[ s\alpha \beta \gamma L_{1}+A(m_{c}^{2}L_{3}-m_{b}^{2}(\beta
+\gamma ))\right] /A^{2},
\end{equation}%
and
\begin{eqnarray}
&&A=\beta \gamma L_{3}+\alpha ^{2}(\beta +\gamma )+\alpha \left[ \beta
(\beta +2\gamma -1)+\gamma (\gamma -1)\right] ,\ B=\alpha ^{2}(\beta -\gamma
)-\gamma L_{3}^{2}+  \notag \\
&&+\alpha (\beta ^{2}-2\gamma (\gamma -1)-\beta (\gamma +1)),\ \ C=\alpha
\beta +\alpha \gamma +\beta \gamma ,\
\end{eqnarray}%
We also use the notations

\begin{equation}
L_{1}=\alpha +\beta +\gamma -1,\ L_{2}=\alpha +\beta -1,~L_{3}=\beta +\gamma
-1,\ L_{4}=\alpha +\gamma -1.
\end{equation}%
The function $\Pi (M^{2},\alpha ,\beta ,\gamma )$ has the following form
\begin{eqnarray}
&&\Pi (M^{2},\alpha ,\beta ,\gamma )=-\frac{\langle \alpha _{s}G^{2}/\pi
\rangle C}{384A^{4}L_{1}\pi ^{4}\beta \gamma }\exp \left[ -\frac{A\left(
m_{b}^{2}(\beta +\gamma )-L_{3}m_{c}^{2}\right) }{M^{2}L_{1}\alpha \beta
\gamma }\right] [m_{b}^{2}(\beta +\gamma )-L_{3}m_{c}^{2}]^{2}  \notag \\
&&(L_{3}^{2}m_{c}^{4}L_{1}\beta \gamma (L_{3}^{2}+3L_{3}\alpha +3\alpha
^{2})+L_{1}m_{b}^{4}\beta \gamma ^{4}(\beta +\gamma
)-m_{b}^{2}m_{c}^{2}\beta \gamma L_{1}(\beta (\beta -1)(2\alpha ^{2}+2\alpha
(L_{2}-\alpha )  \notag \\
&&+(\beta -1)^{2})+\gamma ((\beta -1)^{2}(3\beta -1)+\alpha ^{2}(5\beta
-2)+\alpha (\beta -1)(7\beta -2))+\gamma ^{2}(3-5\alpha +3\alpha ^{2}  \notag
\\
&&+8\beta (\alpha -1)+5\beta ^{2})+\gamma ^{3}(-4+3\alpha +5\beta )+2\gamma
^{4})+m_{b}^{3}m_{c}\beta \gamma (\alpha ^{3}(\beta +\gamma
)^{2}+L_{3}\alpha \gamma (\beta +\gamma )(-3+3\beta +2\gamma )  \notag \\
&&+\alpha ^{2}L_{3}(\beta +\gamma )(\beta +3\gamma )+\gamma L_{3}(\beta
(\beta -1)^{2}+\gamma +\beta \gamma (2\beta -3)+\gamma ^{2}(2\beta -1)))
\notag \\
&&-m_{b}m_{c}^{3}(L_{3}^{2}\alpha \beta (\beta (\beta -1)^{2}+\gamma +\beta
\gamma (6\beta -7)+\gamma ^{2}(9\beta -5)+2\gamma ^{3})+\alpha ^{2}(\beta
^{2}(\alpha ^{2}+2\alpha (L_{2}-\alpha )  \notag \\
&&+2(\beta -1)^{2})(-2+\alpha +2\beta )+\beta (2\alpha ^{3}+7\alpha (\beta
-1)(3\beta -1)+2\alpha ^{2}(5\beta -3)+4(\beta -1)^{2}(5\beta -1))\gamma
\notag \\
&&+(\alpha +\alpha ^{3}+2\alpha ^{2}(4\beta -1)+4\beta (\beta -1)(9\beta
-4)+\alpha \beta (25\beta -18))\gamma ^{2}+(2\alpha (\alpha -1)+\beta
(11\alpha -19)  \notag \\
&&+27\beta ^{2})\gamma ^{3}+\gamma ^{4}(\alpha +7\beta ))+L_{3}^{3}\beta
\gamma (-\gamma +\beta (-1+\beta +2\gamma )))).
\end{eqnarray}%
An explicit formula for $\rho ^{\mathrm{Dim4}}(s,\alpha ,\beta ,\gamma )$ is
rather cumbersome, therefore we do not write down it here.

The correlator $\Pi _{1}(\mathbf{M}^{2},\mathbf{s}_{0},q^{2})$ is given by
the formula Eq.\ (\ref{eq:CorrF1}):%
\begin{equation}
\Pi _{1}(\mathbf{M}^{2},\mathbf{s}_{0},q^{2})=\int_{4\mathcal{M}%
^{2}}^{s_{0}}ds\int_{4m_{b}^{2}}^{s_{0}}ds^{\prime }\rho _{1}(s,s^{\prime
},q^{2})e^{-s/M_{1}^{2}-s^{\prime }/M_{2}^{2}}+\mathcal{B}\Pi _{1}^{\mathrm{%
Dim4}}(p^{2},p^{\prime 2},q^{2}).
\end{equation}%
Here, the function $\rho (s,s^{\prime },q^{2})$ is determined by the
expression
\begin{equation}
\rho _{1}(s,s^{\prime },q^{2})=\frac{3m_{b}m_{c}}{32\pi ^{4}}%
\int_{0}^{1}d\alpha \int_{0}^{1-\alpha }d\beta \int_{0}^{1-\alpha -\beta
}d\gamma \frac{\theta (\Delta )}{(\alpha +\gamma )^{2}L_{4}^{2}},
\end{equation}%
where
\begin{equation}
\Delta =-m_{b}^{2}-\frac{1}{L_{4}}\left[ q^{2}\frac{\alpha \gamma }{\alpha
+\gamma }+m_{c}^{2}(\alpha +\gamma )-s^{\prime }\beta L_{1}\right] .
\end{equation}%
The dimension 4 function $\Pi _{1}^{\mathrm{Dim4}}(p^{2},p^{\prime 2},q^{2})$
is given by the formula%
\begin{eqnarray}
&&\Pi _{1}^{\mathrm{Dim4}}(p^{2},p^{\prime 2},q^{2})=\frac{\langle \alpha
_{s}G^{2}/\pi \rangle m_{b}m_{c}}{192\pi ^{2}}\int_{0}^{1}d\alpha
\int_{0}^{1-\alpha }d\beta \int_{0}^{1-\alpha -\beta }\frac{d\gamma }{\Delta
^{3}L_{4}^{8}(\alpha +\gamma )^{5}}\left\{ -\Delta L_{4}^{2}(\alpha +\gamma
)^{2}\right.  \notag \\
&&\times \left[ 2\alpha ^{4}+6\beta ^{4}+\alpha (12\beta ^{3}+30\beta
^{2}(\gamma -1)+27\beta (\gamma -1)^{2}+8(\gamma -1)^{3})+12\beta
^{3}(\gamma -1)+15\beta ^{2}(\gamma -1)^{2}\right.  \notag \\
&&\left. +9\beta (\gamma -1)^{3}+2(\gamma -1)^{4}+3L_{3}\alpha ^{2}(5\beta
+4\gamma -4)+\alpha ^{3}(9\beta +8\gamma -8)\right] +2(\alpha +\gamma
)^{3}L_{1}\beta \lbrack m_{b}^{2}L_{4}^{2}  \notag \\
&&\left. -p^{\prime 2}L_{1}\beta ][2L_{3}\alpha +\alpha ^{2}+2\beta
^{2}+2\beta (\gamma -1)+(\gamma -1)^{2}]+2L_{4}^{3}\alpha \gamma
(q^{2}\alpha \gamma +m_{c}^{2}(\alpha +\gamma )^{2})\right\} .
\end{eqnarray}

\end{widetext}

\end{document}